\documentclass[reprint,superscriptaddress,showpacs,amsmath,amssymb,aps,prl]{revtex4-1}
\usepackage{graphicx}
\usepackage{dcolumn}
\usepackage{bm}
\usepackage{hyperref}
\usepackage{subfigure}
\hypersetup{colorlinks=true, citecolor=blue, urlcolor=blue, linkcolor=blue}
\usepackage{amssymb}
\usepackage{color}
\usepackage{graphicx}
\usepackage{amsmath}
\usepackage{mathrsfs}
\usepackage{times}
\usepackage{subeqnarray}
\usepackage{cases}
\usepackage{bm}
\usepackage{epstopdf}
\begin{document}
\title{Probabilistic fair behaviors spark its boost in the Ultimatum Game: the strength of good Samaritans}
\author{Guozhong Zheng}
\affiliation{School of Physics and Information Technology, Shaanxi Normal University, Xi'an 710061, P. R. China}
\author{Jiqiang Zhang}
\affiliation{School of Physics and Electronic-Electrical Engineering, Ningxia University, Yinchuan 750021, P. R. China}
\author{Rizhou Liang}
\affiliation{School of Physics and Information Technology, Shaanxi Normal University, Xi'an 710061, P. R. China}
\affiliation{School of Systems Science, Beijing Normal University, Beijing 100878, P. R. China}
\author{Lin Ma}
\affiliation{School of Physics and Information Technology, Shaanxi Normal University, Xi'an 710061, P. R. China}
\author{Li Chen}
\email[Email address: ]{chenl@snnu.edu.cn}
\affiliation{School of Physics and Information Technology, Shaanxi Normal University, Xi'an 710061, P. R. China}

\begin{abstract}
Behavioral experiments on the Ultimatum Game have shown that we human beings have remarkable preference in fair play, contradicting the predictions by the game theory. Most of the existing models seeking for explanations, however, strictly follow the assumption of \emph{Homo economicus} in orthodox Economics that people are self-interested and fully rational to maximize their earnings. Here we relax this assumption by allowing that people probabilistically choose to be ``good Samaritans", acting as fair players from time to time. For well-mixed and homogeneously structured populations, we numerically show that as this probability increases the level of fairness undergoes from the low scenario abruptly to the full fairness state, where occasional fair behaviors ($\sim5\%$) are sufficient to drive the whole population to behave in the half-half split manner.
We also develop a mean-field theory, which correctly reproduces the first-order phase transition and points out that the bistability is an intrinsic property of this game and small fair acts lead to dramatical change due to its bifurcation structure. 
Heterogeneously structured populations, however, display continuous fairness transition; surprisingly, very few hub nodes acting as fair players are able to entrain the whole population to the full fairness state.
Our results thus reveal the unexpected strength of ``good Samaritans", which may constitute a new explanation for the emergence of fairness in our society.
\end{abstract}

\date{\today }
\maketitle
\section{1. Introduction}\label{sec:introduction}
Fairness is one of cornerstones that underpin human civilization, and plays a vital role for solutions to those great challenges we are facing, like climate change, aggravated economic inequality, and pandemics. A canonical model for understanding fairness is the Ultimatum Game (UG)~\cite{Guth1982An}: two players are asked to split a certain amount of money, one (the proposer) proposes a division between herself/himself and the other player (the responder), who either accepts or rejects. If the agreement is reached, they share the money accordingly; otherwise, they earn nothing. 
According to the paradigm of \emph{Homo economicus}~\cite{Simon1957Models,Samuelson2005Economics}, people are self-interested and act in a fully rational manner, the proposers should make the smallest possible offer, because the responder is expected to accept it still because ``something is better than nothing". The rational solution for the UG thus ends with a very unfair split.

A large number of experiments conducted in different countries and with varied incentives, however, reveal that this is not really the scenario how human act when playing the game. The majority of the offers are between $40\%$ and $50\%$ of the total amount, and the offers below $20\%$ are only $3\%$ and with more than half chance are rejected by the responders~\cite{Guth1982An, Thaler1988Anomalies,Bolton1995Anonymity, Roth1995The, Guth2014More}. It turns out that we human are remarkably fond of fairness even at the cost of the potential monetary loss. Similar observations are also made for non-human species, such as monkeys~\cite{Brosnan2003Monkeys} and chimpanzees~\cite{Proctor2013Chimpanzees}, though arousing some debates for the latter~\cite{Jensen2007Chimpanzees}.

Many efforts have been made to understand the above discrepancies. For example, one psychological explanation is that the decision-making for human in the game may be not simply based upon the expected payoff, but also the payoff difference for the two evolved players~\cite{Kirchsteiger1994The, Bethwaite1996The, Fehr1999}. Also, the rejection of a low offer can also be taken as punishment on the proposers, because they lose much more than the responders. In fact, a great variety of theories are proposed in the past thirty years to understand the emergence of fairness~\cite{Debove2016Models}. These theoretical studies indicate that reputation~\cite{Nowak2000Fairness, Andre2011Social}, noise~\cite{Binmore1994An, Gale1995Learning}, empathy~\cite{Sanchez2005Altruism, Page2002Empathy}, population structures~\cite{Page2000The,Kuperman2008The,Sinatra2009The}, and role-alternating~\cite{Iranzo2011The} each could facilitate the evolution of fairness. 
Note that, all these models are developed within the paradigm of \emph{Homo economicus}, players are supposed to be full self-interested and fully rational, all their actions are strictly economic-driven.

This assumption, however, has been challenged by behavioral economists, who have identified the facts that psychological and emotional are important factors in decision-making~\cite{Camerer2003Advances,Loewenstein2003The}, and the related neural basis in the brain are also reported \cite{Sanfey2003The,Rilling2011The}.
In our daily lives, there are some occasions where individuals make their decisions up to some other incentives, such as intrinsic moral incentives; people may occasionally want to a good samaritan to do a good deed without an economic return being expected at all. When they act as such, they are not purely self-interested, and can be regarded as irrational to some extent in the sense that they are now not only economic-driven but also partially other incentive-driven, e.g. morality-driven~\cite{2007The, 2000The}.
It has been argued that our moral sense has developed a sophisticated defense mechanism that enables individuals to survive and the groups they belong to to thrive over millions of years evolution, the morality-driven acts have their place in social activities ~\cite{Boehm2012Moral}.
Thus, it's reasonable to assume the presence of good samaritans, and revealing how they impact the evolution of fair acts would be helpful for understanding the emergence of fairness in many realistic circumstances.

\begin{figure*}[htbp]
\centering
\includegraphics[width=0.4\linewidth]{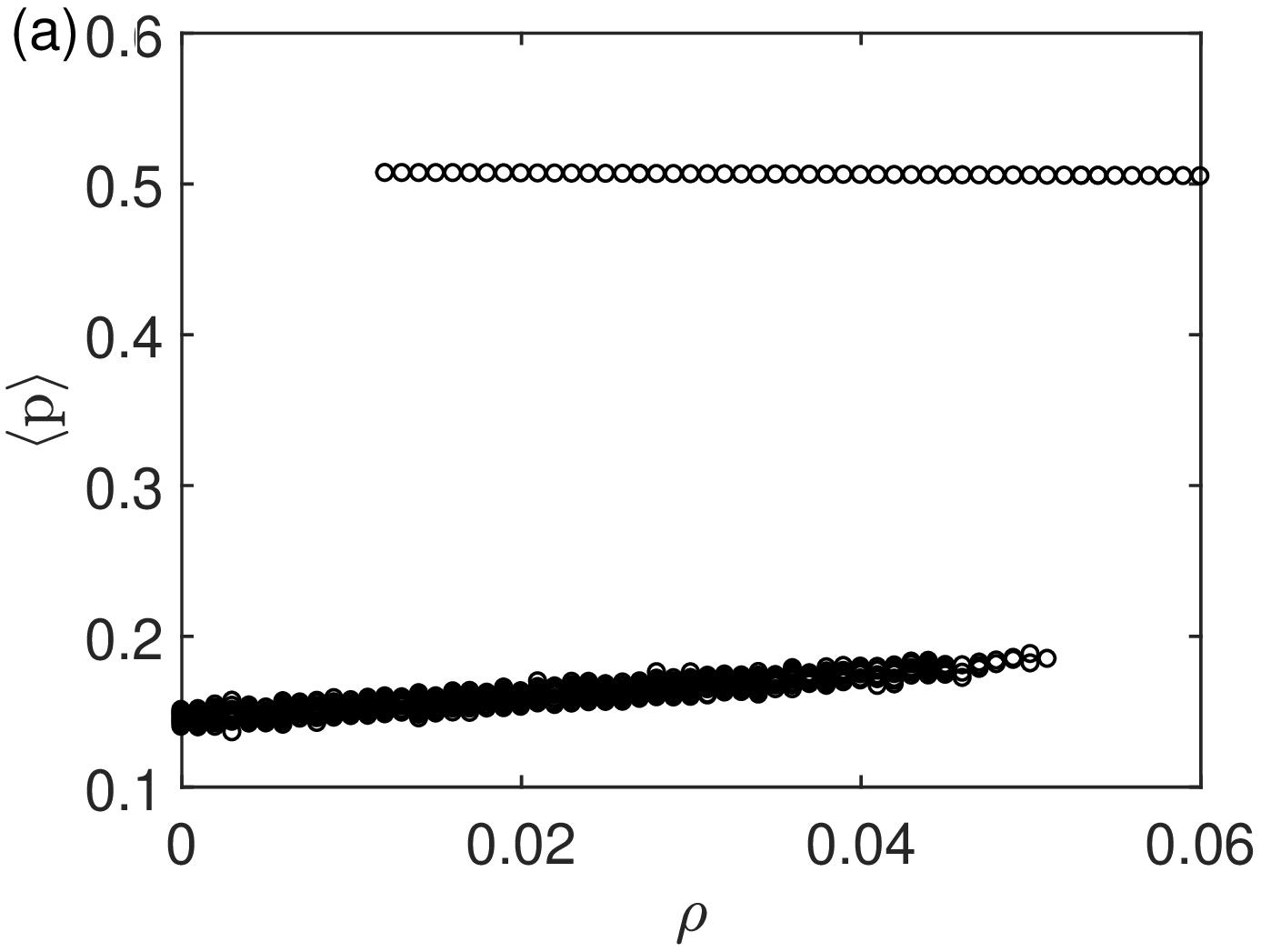}
\includegraphics[width=0.4\linewidth]{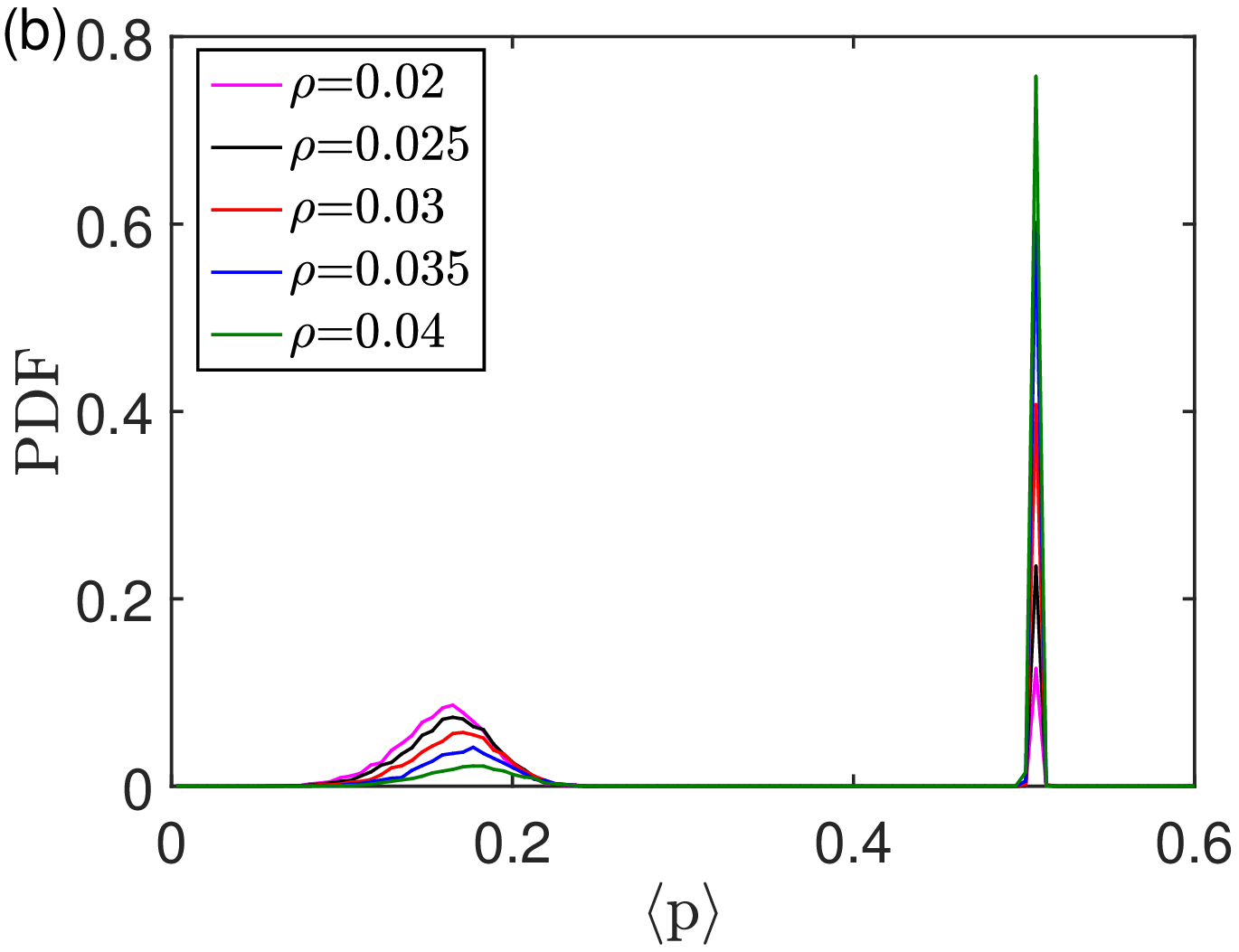}
\includegraphics[width=0.4\linewidth]{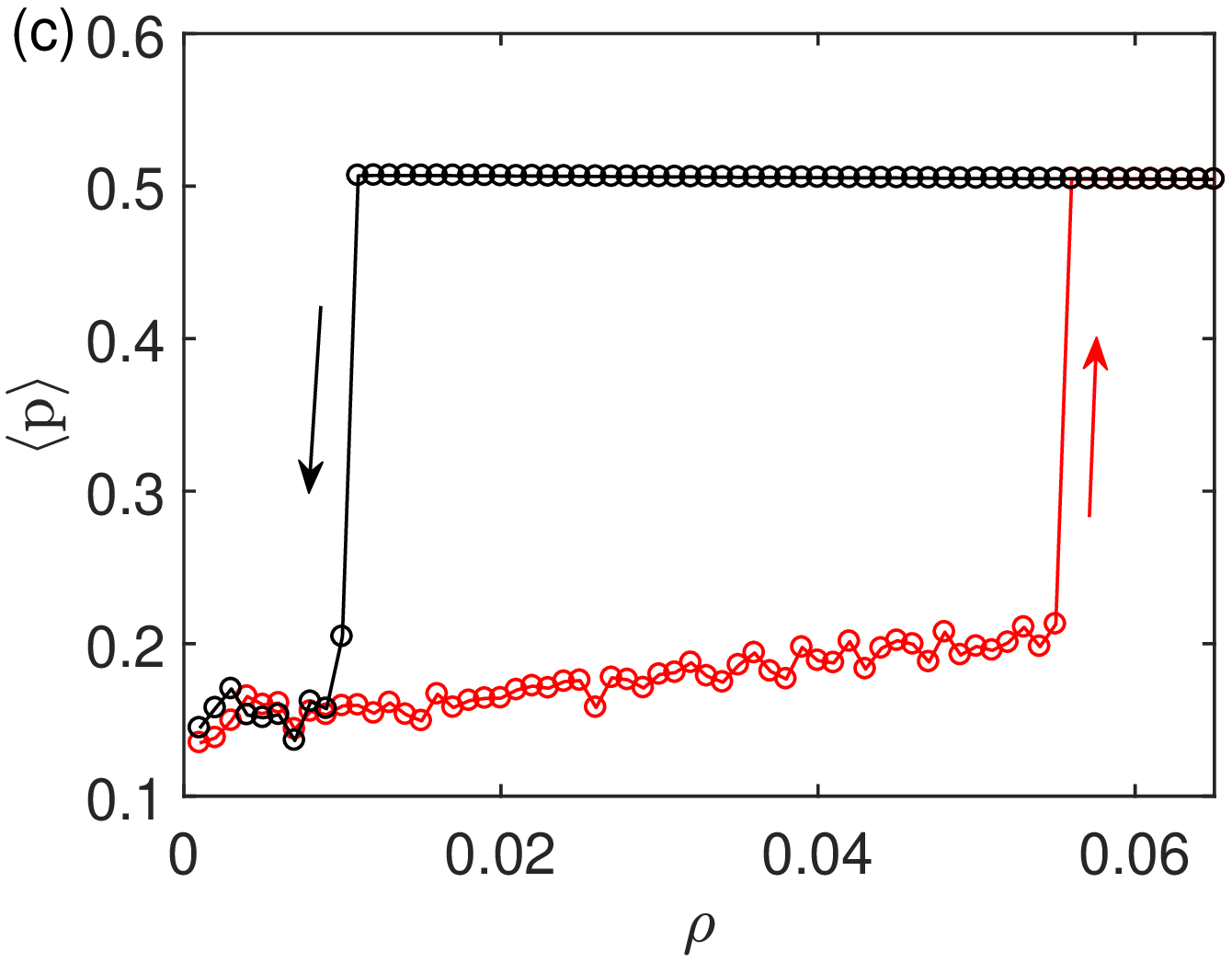}
\includegraphics[width=0.4\linewidth]{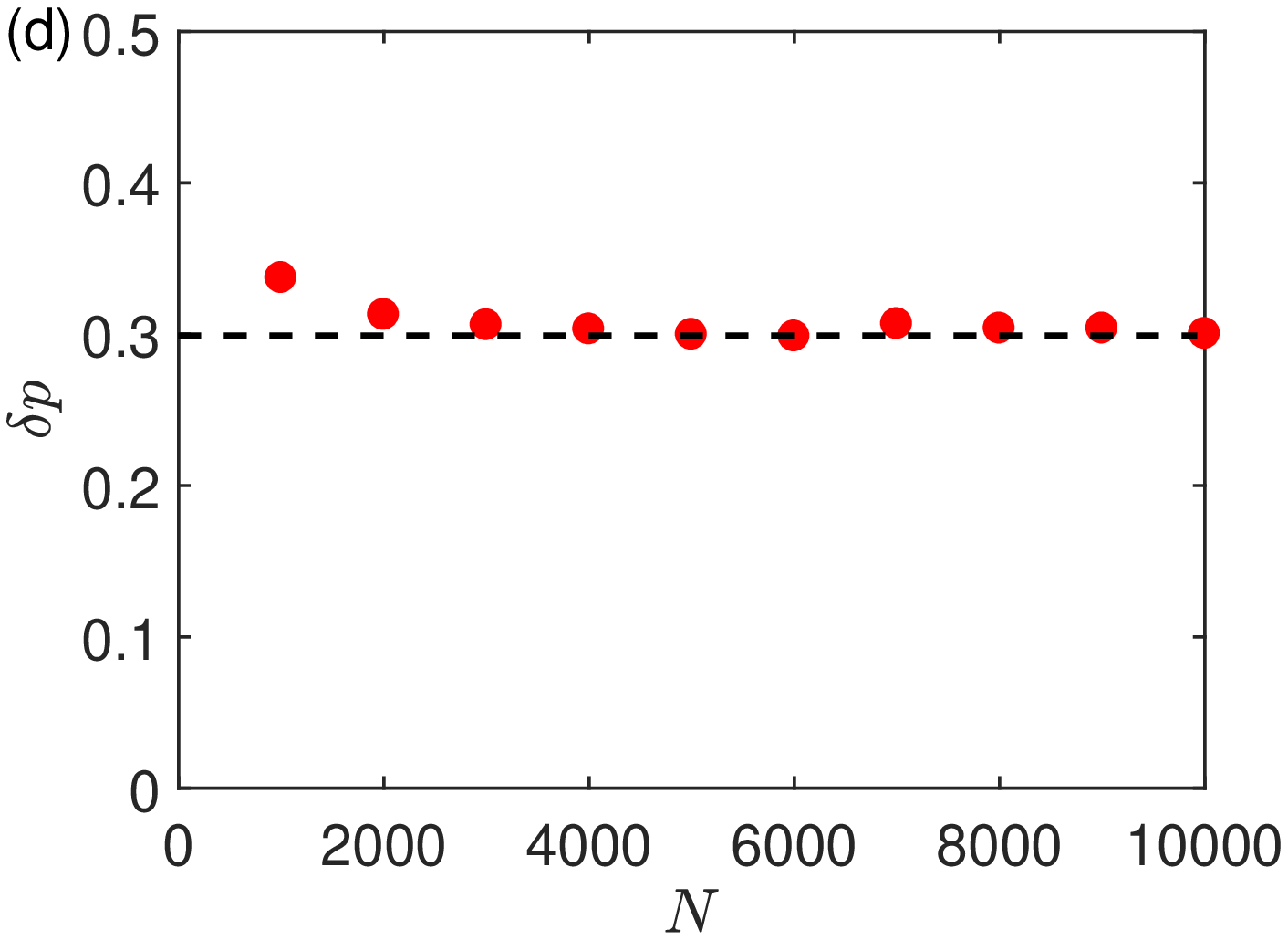}
\caption{(Color online) The evolution of fairness on the $1d$ lattice. 
(a) The phase transitions of the average offer $\langle p\rangle$ versus the spontaneous fair probability $\rho$, each data is averaged $10^6$ times after a transient of $2\times10^6$; it shows an abrupt jump between low fairness states $S_l\approx (0.2,0.2)$ and the high fairness state $S_h\!\approx\!(0.5,0.5)$, i.e. a first-order phase transition.  
(b) The probability density distribution (PDF) curve of $\langle p\rangle$ within the bistable region, the curves present clear bimodal distributions. 
(c) A hysteresis structure is shown with a ramp rate of $1/2\times10^{-8}$ per generation, and each data average over $4\times10^{6}$ time steps. The red and the black lines respectively correspond to the scenarios of increasing and decreasing $\rho$. 
(d) By fixing $\rho=0.03$, the jump size $\delta\rho$ versus the population size $N$, which stabilizes around 0.3 when $N$ becomes large. 
Parameters: $N=1000$ for (a) to (c), and the degree $k=40$.
}
\label{fig:lattice}
\end{figure*}

\begin{figure}[htbp]  
\centering
\includegraphics[width=0.8\linewidth]{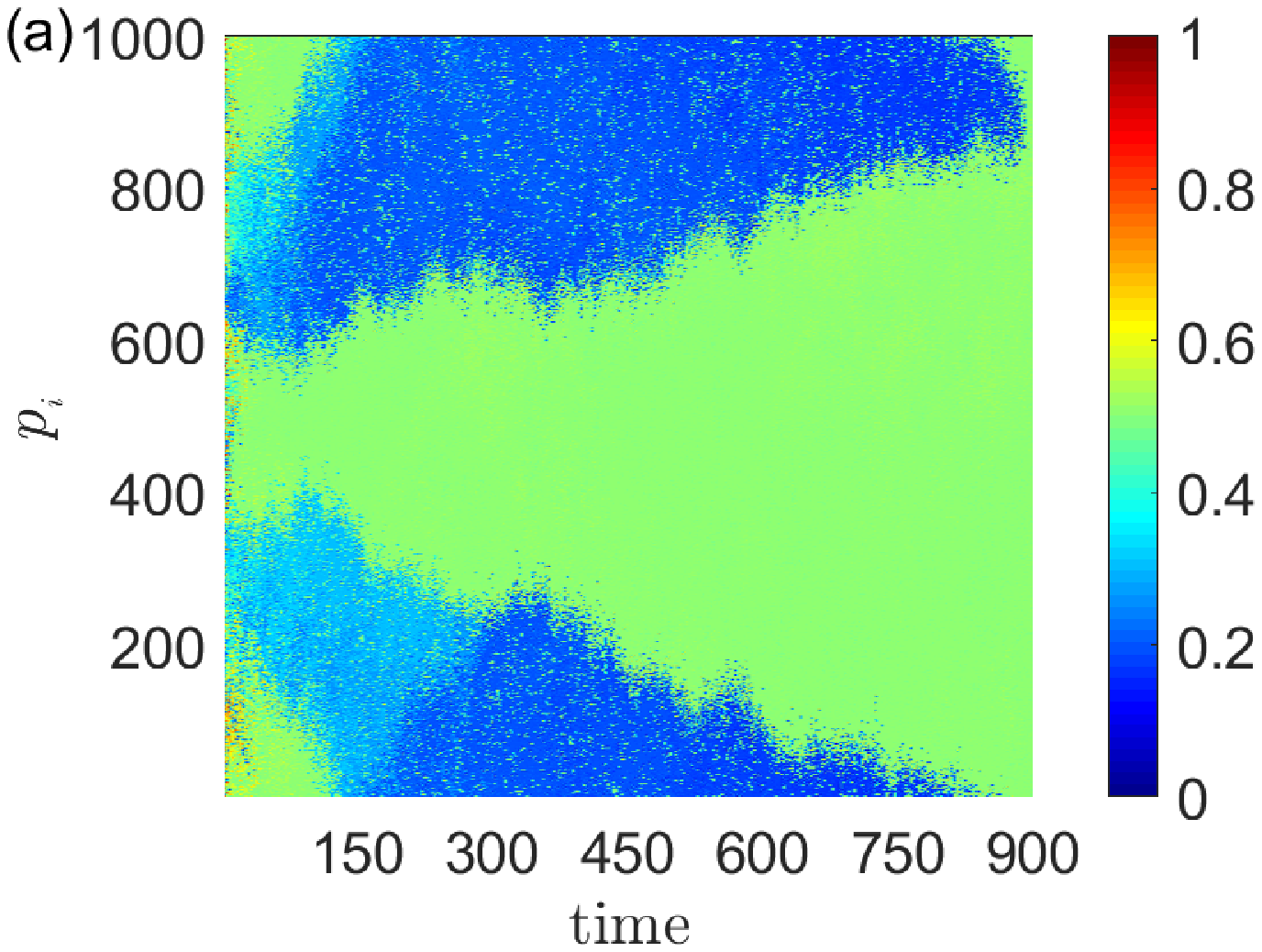}
\includegraphics[width=0.8\linewidth]{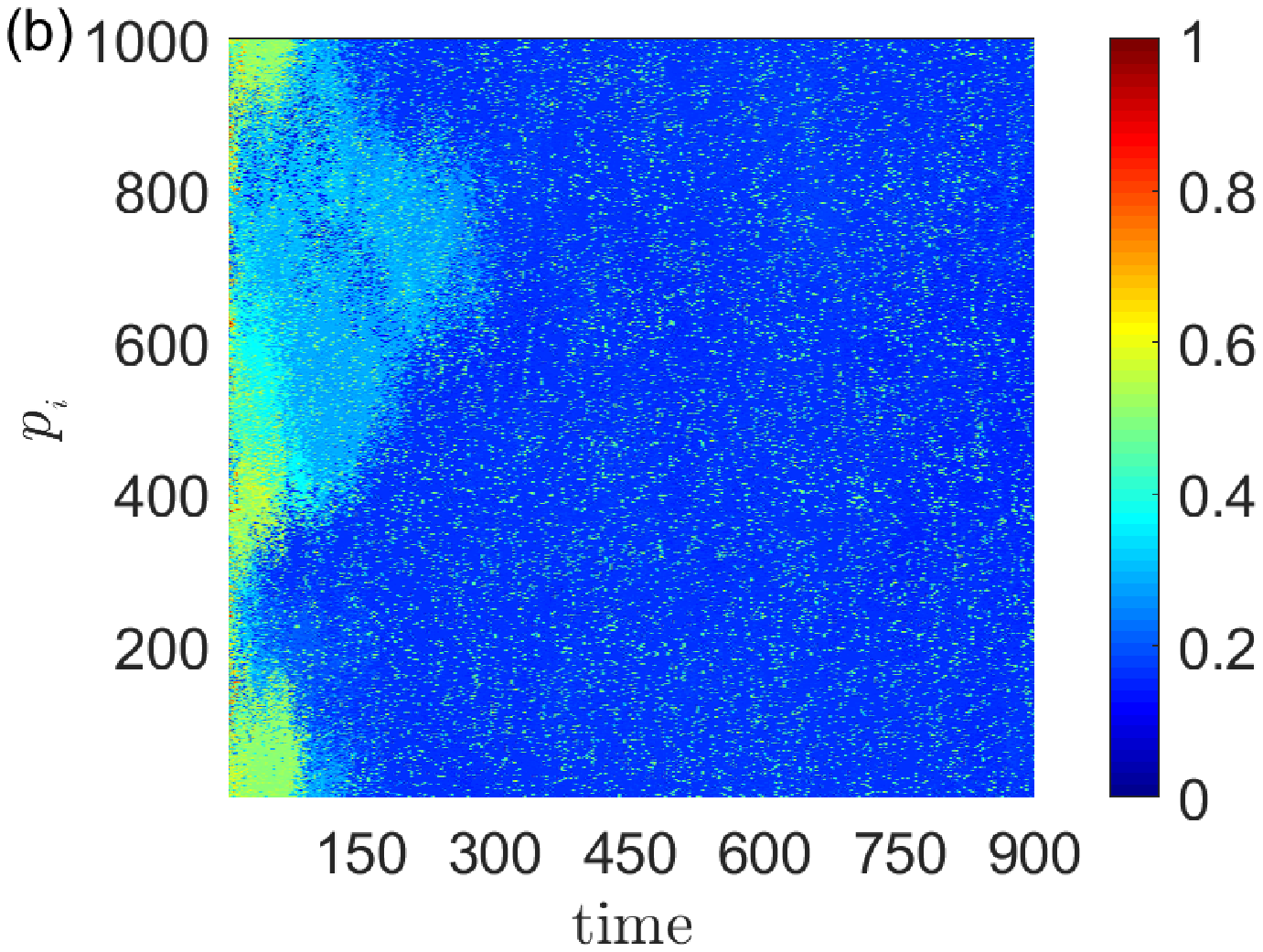}
\caption{(Color online) Spatiotempral evolution of the offer $p_{_i}$ on $1d$ lattice within the bistable region for two different realizations. 
(a) and (b) correspond to scenarios that the high fairness state $S_h$ is successfully and unsuccessfully reached, respectively.
Parameters: $\rho\!=\!0.03$, $N=1000$, and the degree $k=40$.
}
\label{fig:pattern}
\end{figure}

In this work, we fill this gap from the perspective of evolutionary game theory 
~\cite{Kuperman2008The}. By relaxing the assumption of \emph{Homo economicus}, players are allowed to probabilistically act as a fair player from time to time, we investigate the impact of these behaviors on the economic-driven behaviors and the fairness level for the whole population. By tuning the probability, we show that a small such probability is sufficient to drive the whole population to be fair players, an even lower probability is able to sustain this state once it is reached. These form a first-order phase transition of the fairness level and a hysteresis structure is seen. Further analytical treatment and numerical simulations show that these observations are robust to whether the population is unstructured or homogeneously structured, or with different updating rules. Heterogeneous networked population, is further found to enhance the leverage effect, especially when the hub nodes act as fair players; the transitions, however, become continuous.

The paper is organized as follows: 
we introduce our model in Sec. 2\ref{sec:model}. 
In Sec. 3\ref{sec:results}, we show results for the population with 1d lattice and well-mixed structures. 
In Sec. 4\ref{sec:theory}, A mean-field theory is developed for explanation and insights. 
Erd\H{o}s-R\'enyi random networks and Barab\'asi-Albert scale-free networks are also examined to study the impact of underlying population structures in Sec. 5\ref{sec:networks}. Finally, we conclude our work together with some discussions in Sec. 6\ref{sec:discussion}.

\section{2. Model}\label{sec:model}
Let's consider a population of $N$ individuals playing the Ultimatum Game, where they could either be well-mixed or structured (square lattice, random networks, and scale-free networks) in our studies.  Following the common practice~\cite{Guth1982An,Debove2016Models}, the total amount of money to be divided by the two players in each game is equal to the unity. The strategy for a player $i$ is characterized by two parameters $p_i,q_i \in[0,1]$, where $p_i$ is the \emph{offer} that the player $i$ is willing to give to the responder when acting as a proposer, while $q_i$ denotes the \emph{acceptance threshold} when in the role of responder. Their initial values are chosen randomly and independently from the uniform distribution $[0,1]$. 

Suppose now two players $i,j$ are now engaged in an interaction, with their strategies $S_i\!=\!(p_i,q_i)$ and $S_j\!=\!(p_j,q_j)$, respectively. When they are in the role of proposer and responder with equal chance, the expected payoff for player $i$ versus player $j$ is 
\begin{equation}
  E_{ij}= \left\{
   \begin{array}{lr}
	1-p_i+p_j,\quad\quad \mbox{if}\ p_i\geqslant q_j\ \mbox{and}\  p_j\geqslant q_i\\
	1-p_i,\quad\quad\quad\quad\, \mbox{if}\ p_i\geqslant q_j\ \mbox{and}\  p_j< q_i\\
	p_j,\quad\quad\quad\quad\quad\;\;\:  \mbox{if}\  p_i<q_j\:\mbox{and}\  p_j\geqslant q_i\\
	0,\quad\quad\quad\quad\quad\;\;\:\:\,    \mbox{if}\  p_i<q_j\ \mbox{and}\  p_j<q_i.\\
  \end{array}
   \right.
   \label{eq:rule}
\end{equation} 
	
In our numerical experiments, the evolution is as following synchronous updating procedure. At the beginning of each round, with the \emph{spontaneous fair probability} $\rho$, each player chooses to behave as a good Samaritan with the fair strategy $S_h\!\equiv\!(p_h,q_h)$, with $p_h\!=\!q_h\!=\!C_h$ and $C_h=0.5$ here, the half-half split; otherwise it uses the strategy obtained in the last round. 
Next, every player interacts with all their nearest-neighbors up to the underlying networks, their payoffs are collected according to the rule given by Eq.(\ref{eq:rule}) and are then added up, denoted as $\pi_i$ for player $i$. To this end, all players update their strategies proportional to their payoffs, including those adopting the strategy $S_h$, i.e., the player $i$ imitates the strategies in its neighborhood, including itself, according to the Moran rule with the probability~\cite{Roca2009Evolutionary}
\begin{equation}
	w(S_j\rightarrow S_i)=\frac{\pi_j}{\sum_{j\in\Omega'_i}\pi_j}, \label{eq:moran}
\end{equation}
where $\Omega'_i=\Omega_i \cup \{i\}$, and $\Omega_i$ denotes the neighborhood of player $i$. The strategy imitation is subject to small mutations ($\delta p, \delta q)\in[-\varepsilon, \varepsilon]$. Notice that, according to the above rules, the choice of being a good Samaritan is only temporal, no memory is considered; the subsequent action taken is based upon Eq.(\ref{eq:moran}), which can be considered as economic-driven and rational.

Obviously, when the spontaneous fair probability $\rho=0$, our model is recovered to the classic UG within the strict economic-driven paradigm, while the other extreme case of $\rho=1$ is completely mortality-driven, fully fair scenario is expected but is rarely seen. Realistic scenarios are supposed to occur in between the two extremes $0\!<\!\rho\!<\!1$, where both incentives work together. In order to measure the fairness level of the population, we calculate the time averages of offer $\langle p\rangle$ and acceptance threshold $\langle q\rangle$ as the order parameters. A ideal fairness scenario corresponds to  $\langle p\rangle\!=\!\langle q\rangle\!=\!0.5$, the half-half split; whereas the solution for the rational economic man approaches zero if no any other mechanism works. Note that, after the transient evolution, the value of $q_i$ is almost always slightly smaller than $p_i$ for each individual to maximize its payoff, therefore $\langle q\rangle \approx \langle p\rangle$, the so called empathetic scenario~\cite{Sanchez2005Altruism, Page2002Empathy}, and we thus only focus on the evolution of the offer $\langle p\rangle$ to avoid redundancy. 
Finally, if not stated otherwise, the population size $N=1000$, and the mutation strength $\varepsilon=0.001$.

\section{3. Results}\label{sec:results}
\subsection{3.1. Lattices}
We first report the one-dimensional lattice case (see Fig.~\ref{fig:lattice}), where each player connects 20 nearest-neighbors at both sides, therefore the degree $k\!=\!40$. In the absence of spontaneous fair behavior ($\rho\!=\!0$), on average players only offer about $15\%$ share to their partners, for the given parameters, see Fig.~\ref{fig:lattice}(a). Thus the population is in a low fairness state denoted as $S_l\!\equiv\!(p_l,q_l)$, as also found in previous studies~\cite{Kuperman2008The}. As $\rho$ is increased but smaller than $\rho_{c1}\approx0.01$, the overall fairness only improves slightly. Further increasing $\rho$, however, leads to a dramatically different outcome --- the whole population could evolve into a high fairness state $S_h$ for some numerical experiments, yet for some other realizations, the population stays in the low fairness state $S_l$. Finally, when $\rho>\rho_{c2}\approx0.05$, all simulations settle down into the high fairness state, where everyone adopts the fair split strategy $S_h$.

The observation of bistable state is strengthened by the bimodal probability density function (PDF) as shown in Fig.~\ref{fig:lattice}(b). The peak at small values of $p_i$ corresponds to low fairness state $S_l$, and the other to the high fairness state $S_h$ centered around the value of 0.5. As expected, the profile of the low fairness state shrinks when the probability $\rho$ is increased, while the peak of the high fairness state goes up. These features indicate that this is a first-order phase transition (PT) for the fairness evolution in terms of statistical physics.  Another hallmark of first-order PTs is the presence of the hysteresis (see Fig.~\ref{fig:lattice}(c)), where the level of the fairness for a given $\rho$ depends on the history. Importantly, it shows that once the high fairness state $S_h$ is reached, a much smaller probability $\rho\gtrsim 0.01$ is enough to sustain this state. Finally, to examine whether these jumps in the level of fairness are due to the finite-size effect, we check the jump size $\delta p=p_h-p_l$ by increasing the population size $N$ at $\rho=0.03$, where $p_{h,l}$ are respectively the fairness levels at high and low branch in Fig.~\ref{fig:lattice}(c).  Fig.~\ref{fig:lattice}(d) show that the jump size stabilizes around a fixed value 0.3, meaning a reliable first-order PT than the finite-size effect of a continuous PT~\cite{Fan2020Universal}.     

To understand how the spontaneous fair acts induce the unexpected fairness boost for the whole population, let's look at two typical patterns of spatiotemporal evolution by fixing $\rho=0.03$, see Fig.~\ref{fig:pattern}. As shown in Fig.~\ref{fig:pattern}(a), soon after the evolution, a cluster of fair players (i.e. with strategy $S_h$) develops, and this cluster successfully  expands; as time goes by, eventually this fair cluster dominates the whole population. But once there is no such cluster formed at the early stage, the whole population gradually evolves into the low fairness state $S_l$, see Fig.~\ref{fig:pattern}(b). Such observation is analogous to the nucleation phenomena, such as the growth of a small droplet in a supercooled vapor, where the liquefaction depends on the existence of nucleation cores, like dust,  and exhibits a typical first-order phase transition~\cite{1987Introduction}.  

 \begin{figure}[tbp]  
 \centering
\includegraphics[width=0.8\linewidth]{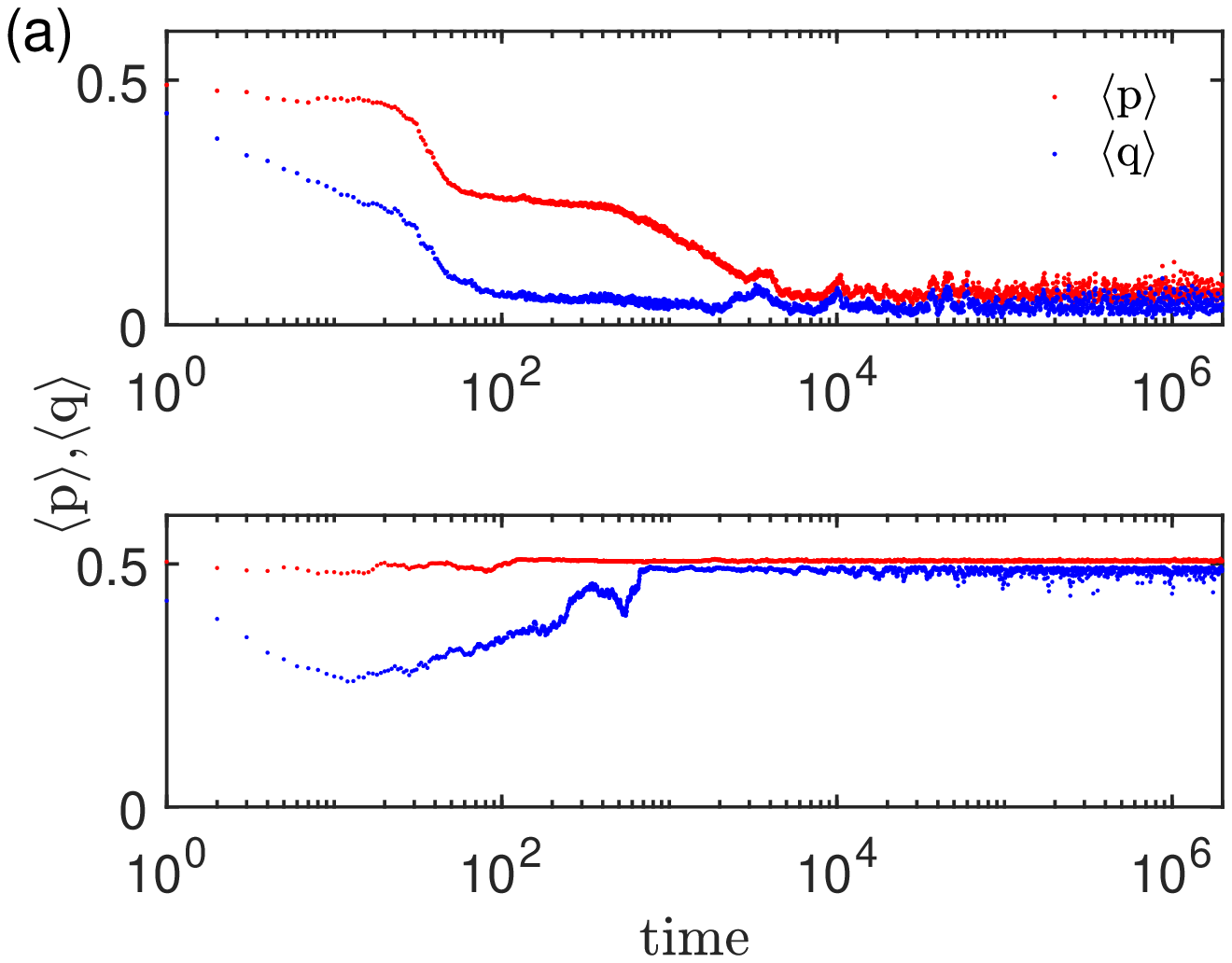}
\includegraphics[width=0.8\linewidth]{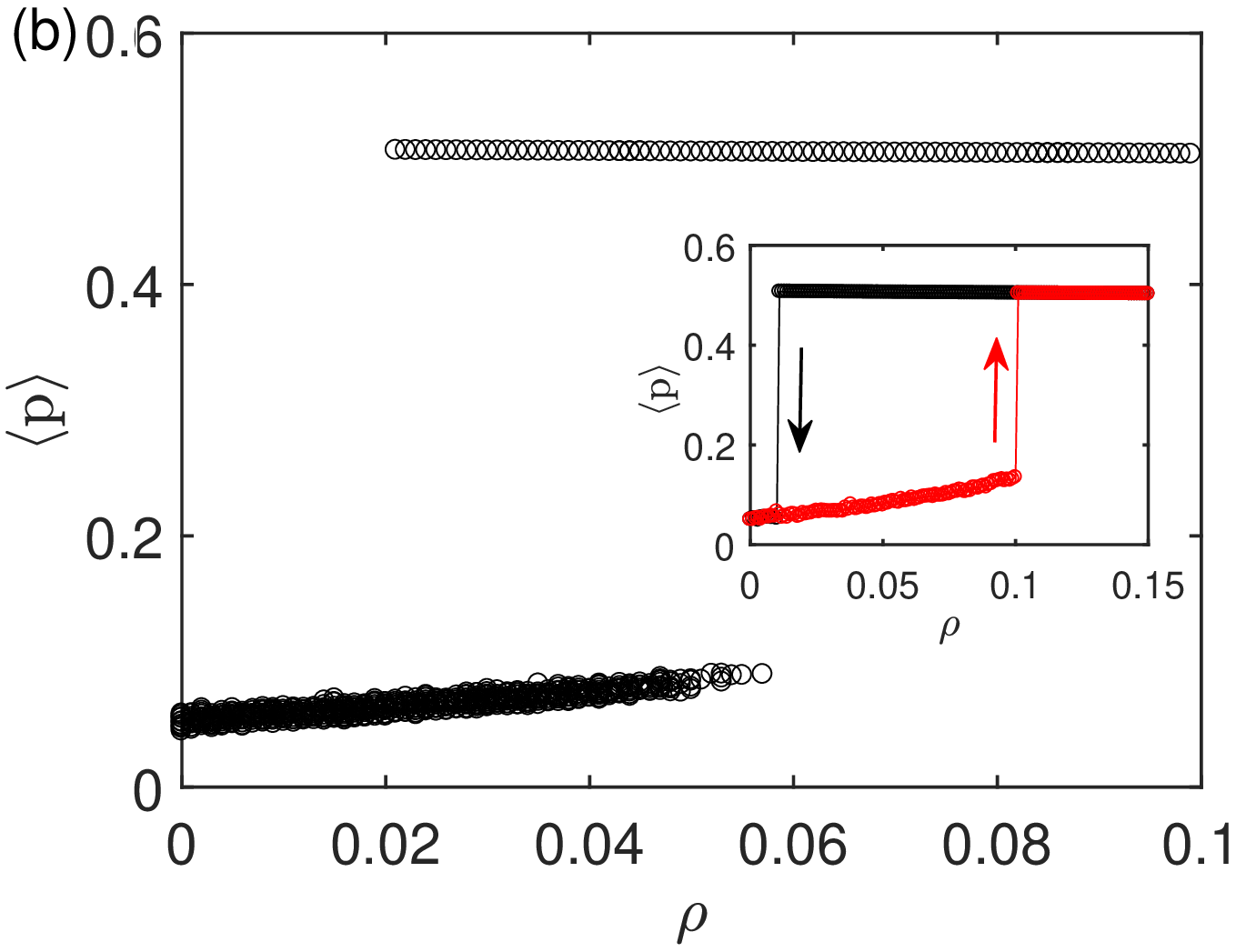}
\caption{
(Color online) The evolution of fairness in the well-mixed population. 
(a) Two time series for the average offer $\langle$p$\rangle$ and acceptance threshold $\langle$q$\rangle$ at $\rho=0.03$, the upper and lower panels correspond to the cases where the population respectively evolves into the low and high fairness states. 
(b) PT of the average offer level $\langle$p$\rangle$ versus the spontaneous fair probability $\rho$, each data is averaged $5\times10^4$ times after a transient $5\times10^4$, where the inset is a hysteresis loop, conducted in the same way as in the above lattice case.
Parameter: $N=1000$.
}
 \label{fig:wellmixed}
 \end{figure}

\subsection{3.2. Well-mixed populations}
Given the fact that the nucleation process is the key to trigger the high fairness state, one might intuitively think the structured nature of population might be necessary for it provides localized circumstances for fair players to get clustered, avoiding the perturbations from the rest part. And since the well-mixed population is unstructured, it appears to be difficult to nucleate, whence the first-order PT seemingly cannot be expected.

In Fig.~\ref{fig:wellmixed}, quite similar observations are made in well-mixed population compared to the above lattice case. Fig.~\ref{fig:wellmixed}(a) shows two time series for $\rho=0.03$, where in the upper panel both the average offer $\langle p\rangle$ and the average acceptance threshold $\langle q\rangle$ decrease to a low level ($\langle p\rangle\!\approx\!\langle q\rangle< 0.1$ here); in the lower panel, however, dramatical rises in both values are seen, where the population are successfully induced to the high level of fairness $S_h$.  This bistability is summarized in the first-order PT shown in Fig.~\ref{fig:wellmixed}(b), where the bistable region is nearly the same as the lattice case (Fig.~\ref{fig:lattice}(a)), but the value of $\langle p\rangle$ in low fairness state is even lower ($p_l<0.1$), due to the absence of network effect~\cite{Page2000The}.  

However, the hysteresis loop shown in the inset of Fig.~\ref{fig:wellmixed}(b) indicates that the bistable region is actually larger, approximately $0.01<\rho<0.1$. Typically, it requires $\rho>10\%$ to entrain the whole population into the fair state $S_h$. The reason for this discrepancy lies in their initial conditions. The initial strategies before jumping up for the inset are $S_l$ for almost all individuals, while the initial $p_i,q_i$ are randomly chosen instead in the main plot, which are then much more likely to nucleate and jump.

These results indicates that the abrupt transition scenario of fairness evolution is robust to whether the population is structured or not.
But, the nucleation process is more likely to occur in structured population, a smaller value of $\rho$ can spark the fair state for the whole population.  

\begin{figure}[tbp]  
\centering
 \includegraphics[width=0.8\linewidth]{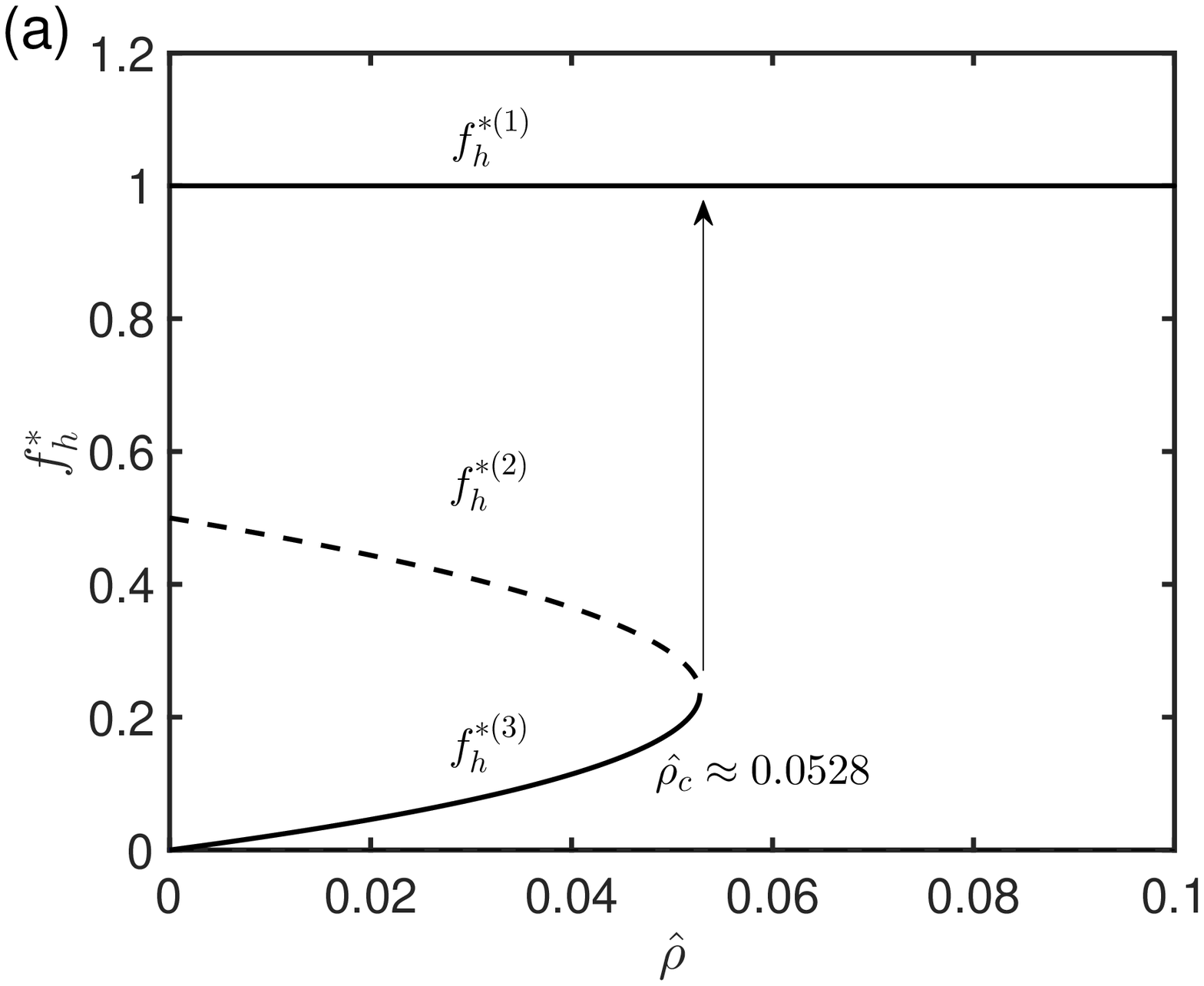}
 \includegraphics[width=0.8\linewidth]{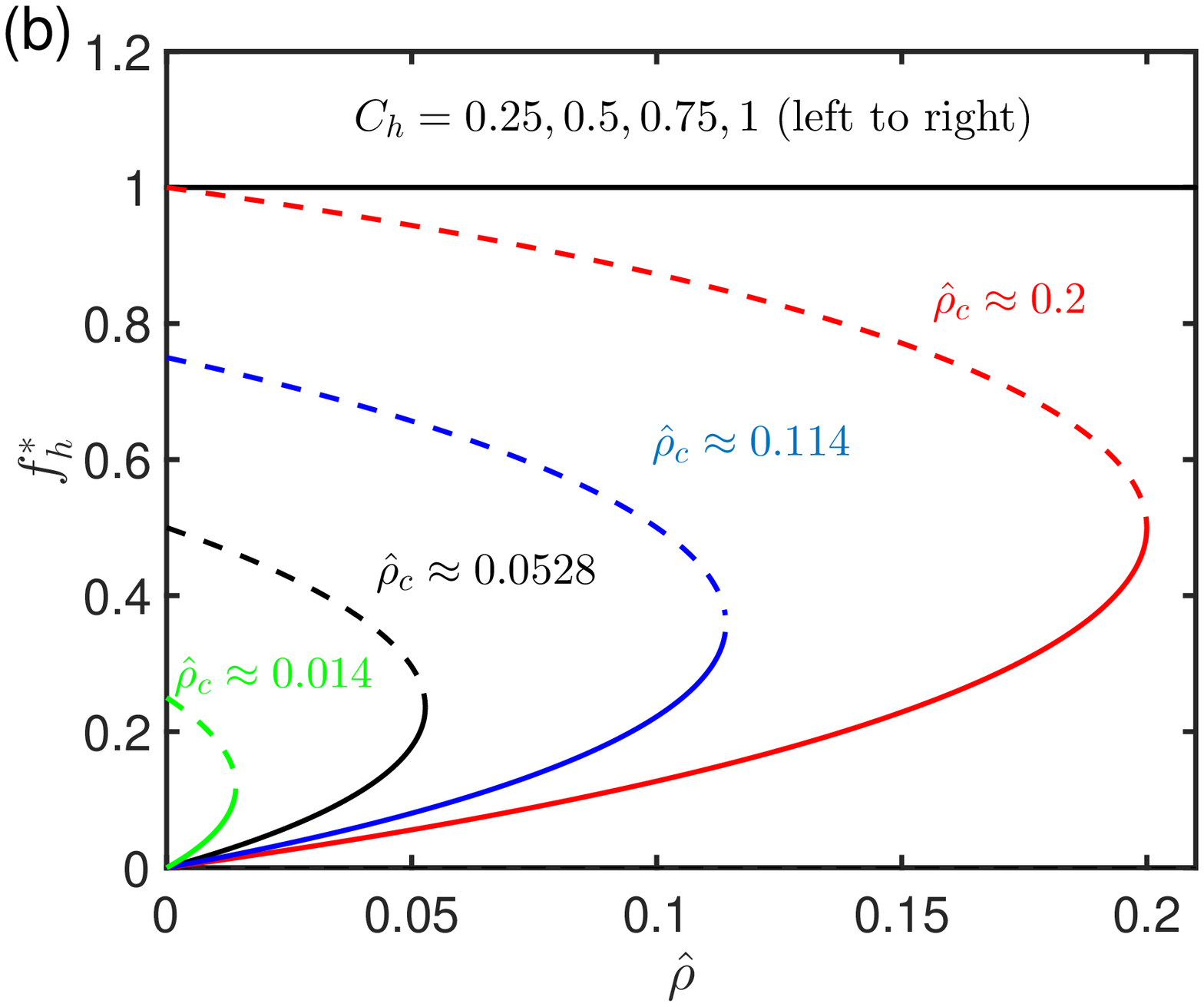}
\caption{(Color online) Results of the mean-field theory.
(a) Bifurcation diagram of high fairness subpopulation fraction $f^*_h$ versus $\hat{\rho}$ when $S_h=(C_h, C_h)$ where $C_h=0.5$. Solid and dashed lines represent stable and unstable fixed solution of the Eq.~(\ref{eq:mf}), respectively. The bistable region is $\hat{\rho}\in[0,\hat{\rho}_c)$. 
(b) Bifurcation diagram for a couple of $C_h$, a higher probability $\hat{\rho}$ is required for a larger value of $C_h$ to reach the state $f^*_h=1$.
}
\label{fig:bifurcation}
\end{figure}

 \section{4. A mean-field theory} \label{sec:theory}
To have a better understanding of the above results, we developed a mean-field theory based on the replicator equation (RE)~\cite{Taylor1978Evolutionary,Roca2009Evolutionary}. Consider an infinitely large well-mixed population, each individual spontaneously chooses the strategy $S_h=(C_h, C_h)$ with rate $\hat{\rho}$ as above. According to the above numerical results, the players typically are either driven to the high fairness state $S_h$, or to a low fairness state  $S_l=(p_l,q_l)$ normally with $p_l\!\approx\!q_l\!<\!0.2$. 

Next, let's denote the fraction for the high and low fairness subgroups are respectively $f_h$ and $f_l$. According to the spirit of RE, the evolution of two fractions 
\begin{equation}
\dot{f}_h=\hat{\rho}_{_0} f_h(\Pi_h-\bar{\Pi})+\hat{\rho}f_l,
\label{eq:fh}
\end{equation}
\begin{equation}
\dot{f}_l=\hat{\rho}_{_0} f_l(\Pi_l-\bar{\Pi})-\hat{\rho}f_l,
\label{eq:fl}
\end{equation}
where $\Pi_{h,l}$ are their payoffs for the two subgroups and $\bar{\Pi}$ is the average payoffs for the whole population. The first term on the right hand side thus describes the selection process due to their relative fitness respective to the whole population with the rate $\hat{\rho}_{_0}$, corresponding to the economic-driven selection. The second term captures the spontaneous transition from the low fairness subgroup $S_l$ to the high fairness subgroup $S_h$ with the rate $\hat{\rho}$, corresponding to the morality-driven process. 
Note that, what really matters here for the competition of the two processes is their relative frequencies, therefore we set $\hat{\rho}_{_0}=1$ without loss of generality and vary the spontaneous fair rate $\hat{\rho}$.
Since $f_h+f_l=1$ has to be satisfied, we shall only focus on Eq. (\ref{eq:fh}). With some algebra (see Appendix), we obtain the evolutionary equation for $f_h$,
\begin{equation}
\dot{f}_h=(1-f_h)[(1-\hat{\rho})f^2_h+(\hat{\rho}-C_h)f_h+ \hat{\rho}].
\label{eq:mf}
\end{equation}
There are three fixed points 
\begin{equation}
\dot{f}^{*(1,2,3)}_h=1,\frac{C_h-\hat{\rho} \pm \sqrt{(\hat{\rho}-C_h)^2-4\hat{\rho}(1-\hat{\rho})}}{2(1-\hat{\rho})}.
\label{eq:mf_solution}
\end{equation}
The bifurcation diagram for $C_h=0.5$ is shown in Fig.~\ref{fig:bifurcation}(a), where $f^{*(1)}_h=1$ is alway a stable solution; one of the other two solutions $f^{*(2)}_h$, however, is unstable. A bistable region is seen when the spontaneous rate $\hat{\rho}<\hat{\rho}_c$, the region terminates via the saddle-node bifurcation~\cite{Kuznetsov2004Elements,Strogatz2015Nonlinear} when the two branches collide, i.e.  $5\hat{\rho}^2-5\hat{\rho}+\frac{1}{4}=0$, leading to $\hat{\rho}_c=\frac{1}{2}-\frac{1}{\sqrt{5}}\approx 0.0528$. Therefore, the predicted bistable region is within $\hat{\rho}\in[0,\hat{\rho}_c)$. 

Importantly, the revealed bifurcation diagram shows that even in the classic UG (i.e. the case of $\hat{\rho}=0$), the dynamics already holds bistable properties, which was pointed out in a previous mini-UG game~\cite{Nowak2000Fairness}. The presence of spontaneous transition process is only to enhance the likelihood for the evolution to the $f^{*(1)}_h$ branch. 
But when $\hat{\rho}\rightarrow 0$, the basin of attraction of the solution $f^{*(1)}_h$ is small that random initial conditions almost always fall into the competing basin for the other stable solution $f^{*(3)}_h$, unless peculiar initial condition being prepared. 
Furthermore, even starting from $f^{*(1)}_h$, the mutation stochasticity could kick the system out of its basin and finally the system evolves into the solution $f^{*(3)}_h$. That's why we see the jump from $S_h$ to $S_l$ when $\hat{\rho}$ is decreased before $\hat{\rho}=0$ being reached in the hysteresis loop. These facts may explain why the bistability nature of UG has long being neglected because the two basins of attraction are not comparable, and thus no related observation can be made. 
As the spontaneous transition rate $\hat{\rho}$ is increased, the basin of attraction of the solution $f^{*(1)}_h$ increases and instead the one of $f^{*(3)}_h$ shrinks. When the two are comparable, bistable properties are easy to see. Finally, when $\hat{\rho}>\hat{\rho}_c$ the basin of $f^{*(3)}_h$ disappears, and $f^{*(1)}_h$ becomes the only stable solution.

The bifurcation diagrams with different $C_h$ are also shown in Fig.~\ref{fig:bifurcation}(b), they show that the bistability nature is robust. The region increases as a high level of fairness is desired, while it disappears as $C_h\rightarrow 0$. This is reasonable because a higher level of fairness requires more effort to spark, e.g. the extreme case of $C_h=1$ requires nearly $20\%$ spontaneous fair acts per round to make the whole population to follow, while only $1.4\%$ is sufficient to guarantee its evolution into the target state $S_h=(0.25,0.25)$.  

A full bifurcation diagram (see Appendix) indicates that the bistable region is actually much larger, the fair society ($f^{*(1)}_h=1$) theoretically can even survive in the case of $\hat{\rho}<0$, which can be interpreted as spontaneous transition from fair state $S_h$ to $S_l$. This prediction is, however, presumptively difficult to observe either in numerical simulations or in real societies, because the basin of attraction of $f^{*(1)}_h=1$ now becomes very small and thus extremely sensitive to any perturbation.

Note that, our mean-field treatment is self-consistent because we assume the existence of the two types of players from very beginning. Its shortcoming is also obvious that it fails to describe the transient dynamics, such as the random initial conditions used in our work. 

\section{5. Random and scale-free networks}\label{sec:networks}
In the real world, the population is structured neither in any regular fashion nor unstructured as studied above, the interpersonal connectivities are often depicted as complex networks~\cite{1998Collective}. Here, we investigate Erd\H{o}s-R\'enyi (ER) random networks~\cite{Bollobas2001random} and Barab\'asi-Albert (BA) scale-free networks~\cite{Barabasi1999Emergence}, which represent typical cases of homogeneous and heterogeneous complex networks. We examine their impact on the outcome of our model. For comparison, all network sizes are also $N$=1000, and the average degree $\langle k\rangle=4$.

\begin{figure}[htbp]  
\centering
\includegraphics[width=0.8\linewidth]{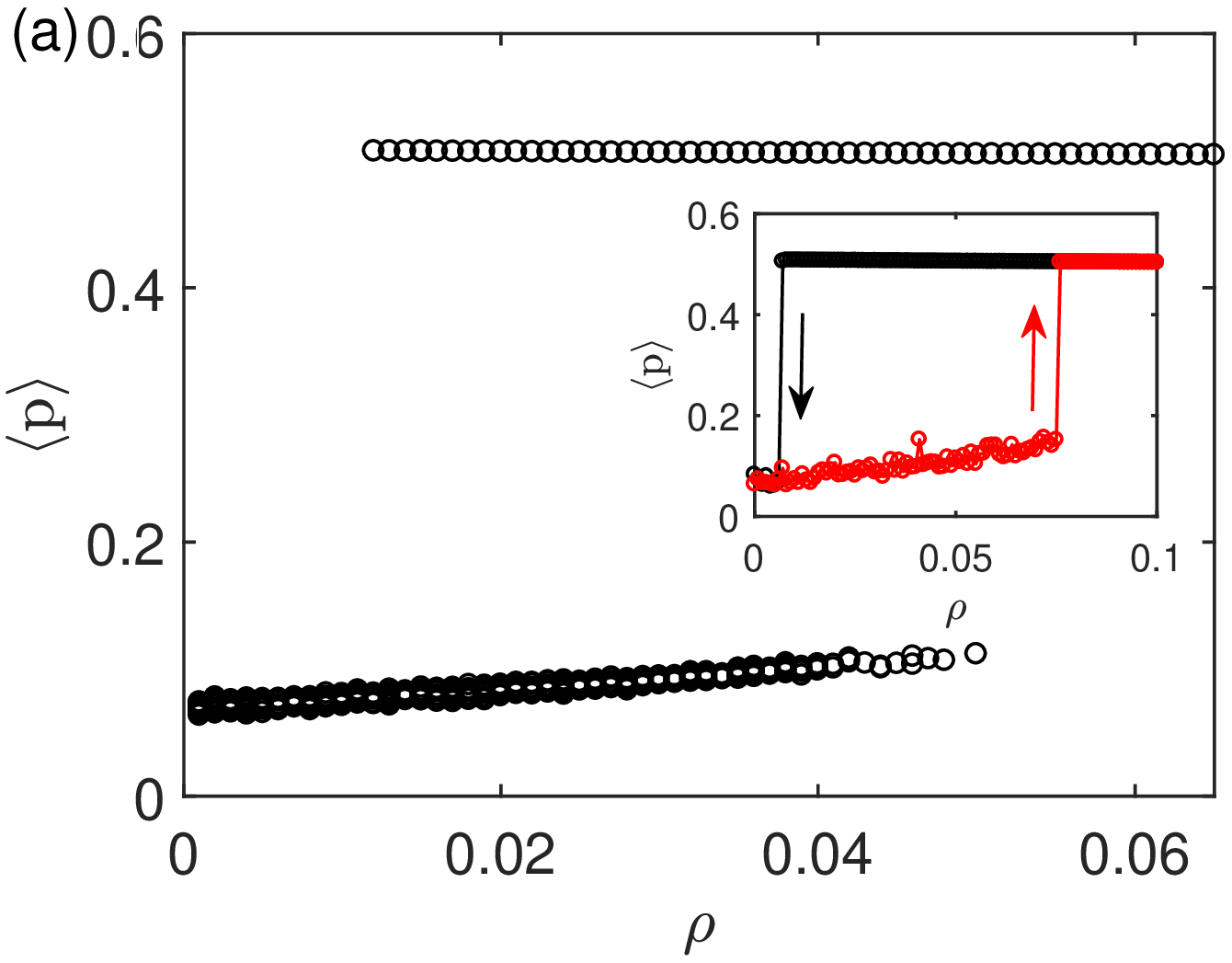}
\includegraphics[width=0.8\linewidth]{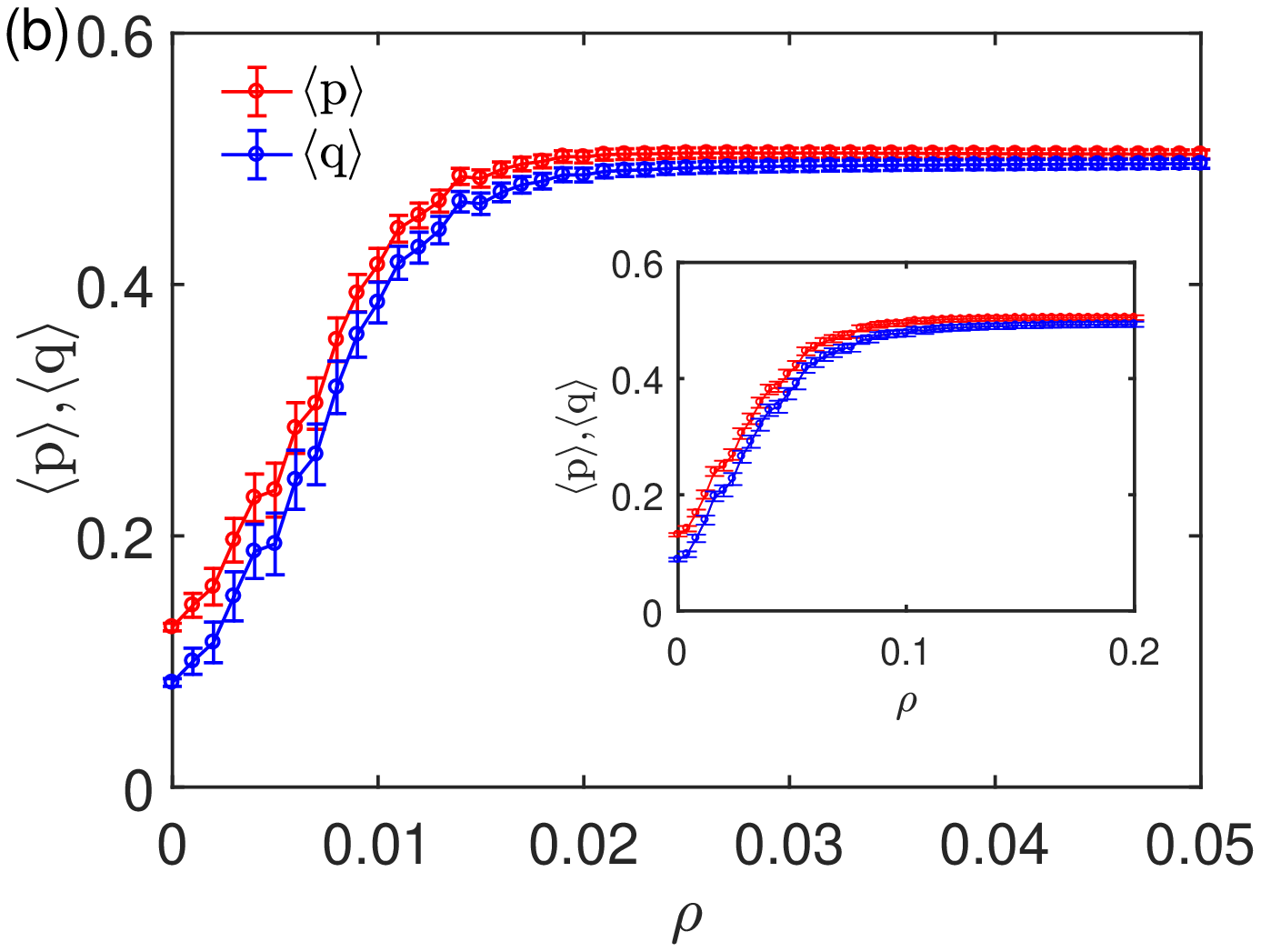}
\caption{(Color online) The fairness level versus the probability $\rho$ on ER random networks (a) and BA scale-free networks (b). 
(a) A first-order PT of the average fairness $\langle$p$\rangle$  on ER networks is also seen, where each data is averaged $10^6$ times after a transient of $10^6$. A hysteresis is also shown in the inset with the same setup as in the lattice case.
(b) Continuous transitions of $\langle$p$\rangle$ and $\langle q\rangle$ are shown instead in BA networks, the inset shows the case where only the node with the largest degree probabilistically acts in a fair manner versus the probability $\rho$, the rest are purely economic-driven. Each data is averaged 20 realizations, and for $2\times10^6$ time average after a transient of $8\times10^6$. The error bars represent their standard deviations.  
Parameters: $N$=1000 and $\langle k\rangle=4$. 
}
\label{fig:ERSF}
\end{figure}

Fig.~\ref{fig:ERSF}(a) reports the case of ER random networks, where similar observations are made. The bistable region within the first-order fairness transition $0.01<\rho<0.05$ is almost the same compared to the above two cases. The hysteresis in the inset, indicates the $\rho_{c2}$ is also actually larger, around 0.075, due to the same reason given in the well-mixed population, but with a lesser extent.
This then provides evidence that the boost of the overall fairness level by probabilistic fair behaviors is robust for homogenous structured populations. 

For BA scale-free networks, however, the transition from low to high level of fairness is no longer in a discontinuous manner, see Fig.~\ref{fig:ERSF}(b). In the absence of spontaneous fair act $\rho=0$, the average strategy $(\langle p\rangle, \langle q\rangle)\approx (0.12, 0.1)$.
As the probabilistic fair acts set in, the value of $\langle p\rangle$ continuously improves as $\rho$ is increased, and full fairness state $S_h$ is reached when $\rho>0.02$, smaller than the typical value of $\rho_{c2}\approx0.05$ in the ER network case. This means that the network heterogeneity further facilitates the boost of fairness sparked by the probabilistic fair behaviors, but in a smooth way. 

Intuitively, on heterogeneous networks, the hub nodes are expected to outperform the peripheral nodes in boosting the fairness level because they can affect much more neighbors. To validate, only one hub with the largest degree is chosen to spontaneously act within $S_h$ with probability $\rho$, the rest are purely economic-driven, evolving according to Eq.(\ref{eq:moran}). Surprisingly, when $\rho>0.2$, the desired fairness for the whole population is reached, see the inset in Fig.~\ref{fig:ERSF}(b). This means, when the hub players spontaneously act in the fair manner, their impact will disproportionally spreads to the whole population that could be unexpectedly to overturn the fairness level.

Heuristically, one might guess the hubs are more easily to form fair clusters that entrain the rest of the network to be in the state $S_h$, just like the synchronization process~\cite{GG2007Paths} or the cooperation evolution~\cite{Santos2005Scale-Free} on the scale-free networks in previous studies. This guess however is not true, the fact is much more complex; the hubs are found that not more likely to form fair clusters, in fact the evolution of fairness is strongly fluctuating when $\rho$ is small, long intermittency of low fair state are found with nontrivial statistical features, which will be presented elsewhere.

\section{6. Discussion}\label{sec:discussion}
In summary, in this work we relax the assumption of \emph{Homo economicus} and allow probabilistic spontaneous fair acts, and focus on their impact on the evolution of the fairness level. We find that as this probability increases, a first-order PT is identified for all cases except for the heterogeneously structured populations. The level of fairness for the population could abruptly jumps from a low fairness state into a high fairness state. Once the population is heterogeneously structured, such as in a scale-free network manner, the transition becomes continuous, and the network heterogeneity is found to further enhance this leverage effect. Further simulations show that these findings are robust against model details, e.g., when the synchronous updating is replaced with asynchronous version~\cite{Roca2009Evolutionary}, or the Moran rule is replaced with the Fermi rule~\cite{Szabo1998Evolutionary}. Our mean-field treatment reveals that there is a bistable fairness structure for the evolution in UG model, and explains why the spontaneous acts make the jump to the high fairness state easier. The implication of our findings are that once occasional spontaneous fair acts are present, the emergence of a high level of fairness is possible. Since probabilistic fair behaviors are indeed often seen in the real life, the observations made here is expected to work as a plausible mechanism for the emergence of fairness.

Interestingly, part of these observations were confirmed in a previous experiment work by Brenner et al~\cite{Brenner2006On}, where the human play together with the artificial players with assigned strategies, they found that a lower acceptance rate or higher $q$ leads to higher offers, i.e. a higher level of fairness. This observation was previously modeled by integrating both fairness motive and adaptive learning hypotheses, and found that the presence of a small fraction of ``tough'' responders who reject unfair offers indeed results in a leverage effect for the whole population~\cite{Xiong2014Emergence}. Though, our work points out the possibility of abrupt jump and the underlying of bifurcation structure, besides the fairness motive alone is sufficient to account for the leverage effect.

From the control point of view, our work may also provide a strategy to promote social fairness in some contexts where the fairness level is not satisfied. Instead of probabilistic fair acts, a fraction of individuals are pinned to be fair players, who are supported financially to keep that strategy, and $\rho$ is interpreted as the fraction of pinned players. Preliminary simulations show that this pinning control is very efficient to achieve a high fairness state, with relatively low control costs, which will be presented elsewhere.  

We note that our model is slightly deviated from the paradigm of \emph{Homo economicus} that spontaneous fair acts are allowed and we focus on its consequence by evolution. The assumption per se, however, is made \emph{a prior}, motivated by the daily experience. Our work cannot explain why occasional fair behaviors are present, which has been an important subject in psychology~\cite{Boehm2012Moral}.

\section{Acknowledgments}
We are supported by the Natural Science Foundation of China under Grants Nos. 61703257 and 11747309, and by the Fundamental Research Funds for the Central Universities GK201903012. ZJQ is supported by the Natural Science Foundation of China under Grants No. 12165014.

\appendix
\section{Appendix: Deviation and analysis of the mean-field treatment}\label{app:A}
Following the notation in Sec. 4, the expected payoffs for the players in the state $S_h$ and $S_l$ respectively are 
\begin{equation}\label{eq:pi1}
\begin{aligned}
\Pi_{h}=[f_h+(1-f_h) \hat{\rho}]\pi_{hh}+ (1-f_h)(1-\hat{\rho})\pi_{hl},
\end{aligned}
\end{equation}
\begin{equation}\label{eq:pi2}
\begin{aligned}
\Pi_{l}= [f_h+(1-f_h) \hat{\rho}]\pi_{lh}+(1-f_h)(1-\hat{\rho})\pi_{ll},
\end{aligned}
\end{equation}
where, e.g. $\pi_{hl}$ denotes the payoff for the player within $S_h$ when she/he encounters a player within $S_l$. According to the model setting, $\pi_{hh}=\pi_{ll}=1$, $\pi_{hl}=1-C_h$, and $\pi_{lh}=C_h$. The mean payoff by definition is
\begin{equation}
\bar{\Pi}= f_h\Pi_h+f_l\Pi_l.
\label{eq:pi_mean}
\end{equation}
With these, we  insert Eq.~(\ref{eq:pi1}-\ref{eq:pi_mean}) in to Eq.~(\ref{eq:fh}), which leads to Eq.~(\ref{eq:mf}), a nonlinear ordinary differential equation.

\begin{figure}[htbp]  
\centering
 \includegraphics[width=0.7\linewidth]{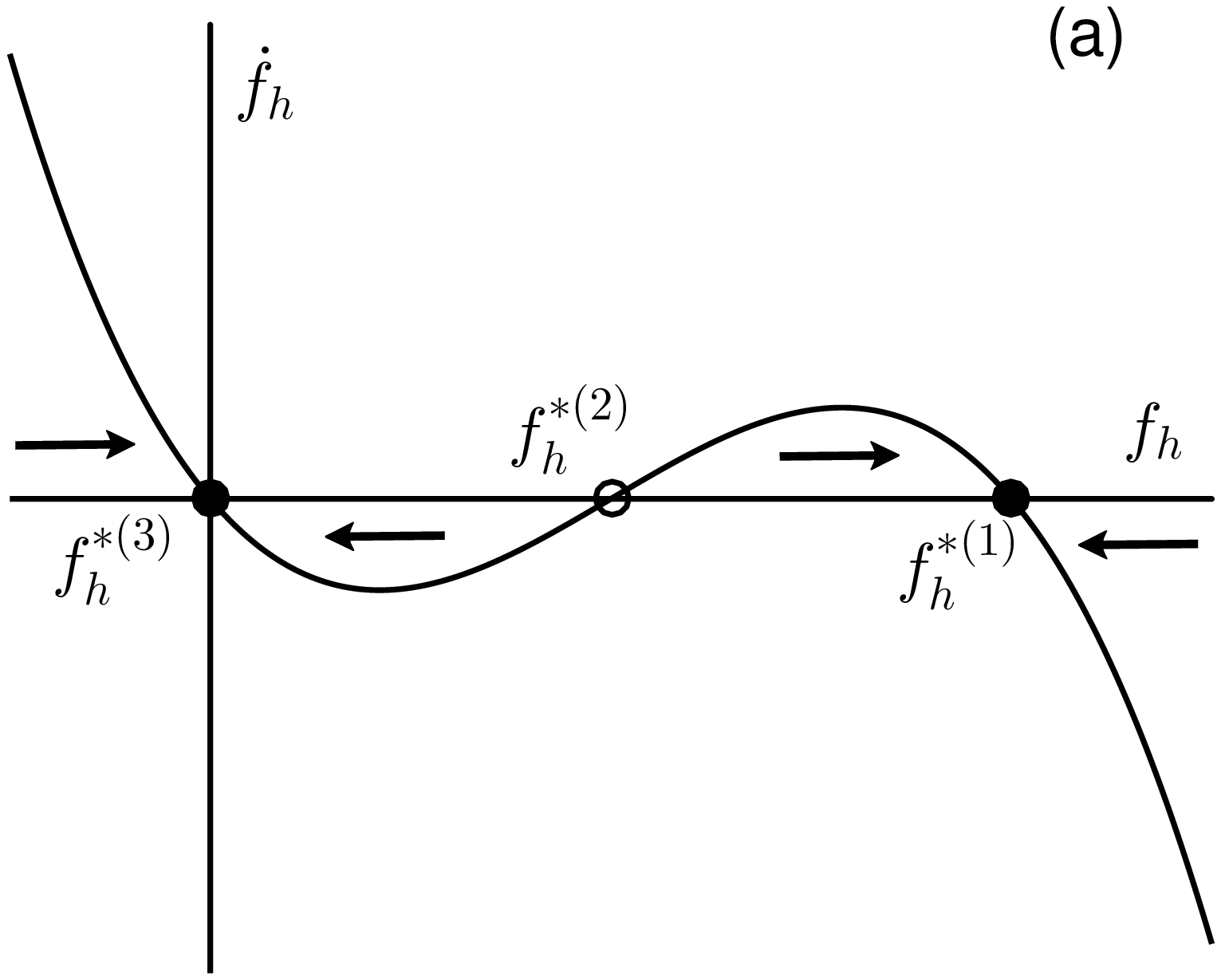}
 \includegraphics[width=0.7\linewidth]{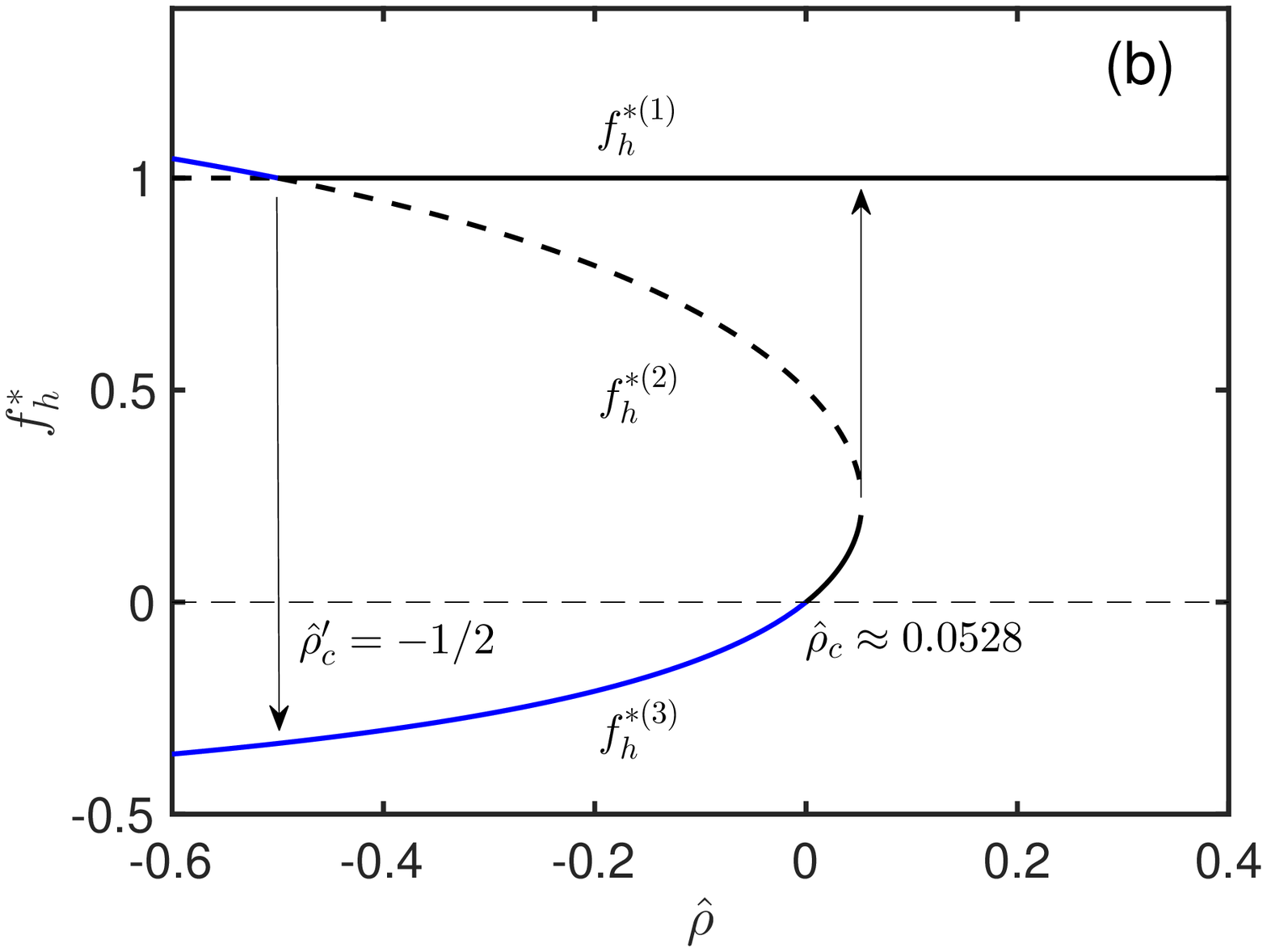}
\caption{(Color online) 
(a) Geometrical illustration of Eq.~(\ref{eq:mf}) within $f_h - \dot{f}_h$ axis for the case of $\hat{\rho}=0$. Three points of intersection corresponds to the fixed points, solid and open circles are respectively of stable and unstable types. Arrows indicate the moving directions determined by the sign of $\dot{f}_h$. 
(b) Full bifurcation diagram compared to Fig.~\ref{fig:bifurcation}(a). Solid and dashed lines are stable and unstable fixed solutions of the Eq.~(\ref{eq:mf}), respectively; the blue lines are unphysical because the value is out of reasonable range. The full bistable region is $\hat{\rho}\in(\hat{\rho}'_c,\hat{\rho}_c)$. Parameter: $C_h=\frac{1}{2}$.
 }
\label{fig:bifurcation_full}
\end{figure}

To better understand the mathematical structure of Eq.~(\ref{eq:mf}), it's helpful to visualize it geometrically~\cite{Strogatz2015Nonlinear}. Fig.~\ref{fig:bifurcation_full}(a) within $f_h - \dot{f}_h$ axis provides such an example for the case of $\hat{\rho}=0$. The fixed points $f^{*(1,2,3)}_h=(1,\frac{1}{2},0)$ as the solution of the equation are the intersection points of the curve of Eq.~(\ref{eq:mf}) where $\dot{f}_h=0$. But due to the stability requirement, only  $f^{*(1)}_h$ and $f^{*(3)}_h$ are stable because only the slopes around these two fixed points are negative. Otherwise, any perturbation around a fixed with a positive slope, e.g. at $f^{*(2)}_h$, will grow and not go back.

When the moral probability $\hat{\rho}$ is increased from zero, the two intersection points $f^{*(2)}_h$ and $f^{*(3)}_h$ move close to each other, as also shown in Fig.~{\ref{fig:bifurcation}}. When $\hat{\rho}=\hat{\rho}_c$, the two points collide with each other, and they disappear when $\hat{\rho}>\hat{\rho}_c$. This is then a saddle-node bifurcation. Thus, $f^{*(1)}_h$ becomes the only stable fixed point when $\hat{\rho}>\hat{\rho}_c$, the whole population are within the state $S_h$. 

Mathematically, it's also intriguing to decrease $\hat{\rho}$ to be negative to obtain the complete bifurcation diagram, as shown in Fig.~\ref{fig:bifurcation_full}(b) for the case $C_h=\frac{1}{2}$. When $\hat{\rho}<0$, this can be interpreted as the scenario that people probabilistically choose the strategy $S_l$, spontaneously being a low fairness player, just opposite to the good Samaritan case studied above where $\hat{\rho}>0$. By doing so, the two intersection points $f^{*(2)}_h$ and $f^{*(3)}_h$ now move far away from each other in Fig.~\ref{fig:bifurcation_full}(a), but then $f^{*(3)}_h$ becomes negative, which is unphysical. Further decreasing $\hat{\rho}$ leads to a transcritical bifurcation~\cite{Kuznetsov2004Elements,Strogatz2015Nonlinear}, where $f^{*(1)}_h$ and $f^{*(2)}_h$ switch their stability, but then the stable solution $f^{*(2)}_h>1$ becomes also unphysical. The complete bistable region is then within $\hat{\rho}\in[\frac{1}{2},0.0528)$, see Fig.~\ref{fig:bifurcation_full}(b).

The interpretation for this region is as follows: 

i) when $\hat{\rho}\in[-\frac{1}{2},0]$, the system is still in a bistable region in both mathematical and physical sense. When the initial condition falls within the basin of attraction of the branch $f^{*(1)}_h$, the evolution of the fairness goes to $f^{*(1)}_h$, though this basin shrinks as $\hat{\rho}$ is decreased. Otherwise, the fairness settles down at $f_h=0$ in the realistic situation, because this is an absorbing state, no further changes would be expected; the solution $f^{*(3)}_h<0$ will not be reached. 

ii) when $\hat{\rho}<-\frac{1}{2}$, the system is bistable only in mathematical sense. Neither the solution $f^{*(2)}_h>1$ nor $f^{*(3)}_h<0$ will be approached, $f_h=0$ is the only stable solution, because any initial condition falls within the basin of attraction of $f^{*(3)}_h$ and the population always approaches the absorbing state $f_h=0$.

\bibliography{fairness}

\begin{thebibliography}{45}%
\makeatletter
\providecommand \@ifxundefined [1]{%
 \@ifx{#1\undefined}
}%
\providecommand \@ifnum [1]{%
 \ifnum #1\expandafter \@firstoftwo
 \else \expandafter \@secondoftwo
 \fi
}%
\providecommand \@ifx [1]{%
 \ifx #1\expandafter \@firstoftwo
 \else \expandafter \@secondoftwo
 \fi
}%
\providecommand \natexlab [1]{#1}%
\providecommand \enquote  [1]{``#1''}%
\providecommand \bibnamefont  [1]{#1}%
\providecommand \bibfnamefont [1]{#1}%
\providecommand \citenamefont [1]{#1}%
\providecommand \href@noop [0]{\@secondoftwo}%
\providecommand \href [0]{\begingroup \@sanitize@url \@href}%
\providecommand \@href[1]{\@@startlink{#1}\@@href}%
\providecommand \@@href[1]{\endgroup#1\@@endlink}%
\providecommand \@sanitize@url [0]{\catcode `\\12\catcode `\$12\catcode
  `\&12\catcode `\#12\catcode `\^12\catcode `\_12\catcode `\%12\relax}%
\providecommand \@@startlink[1]{}%
\providecommand \@@endlink[0]{}%
\providecommand \url  [0]{\begingroup\@sanitize@url \@url }%
\providecommand \@url [1]{\endgroup\@href {#1}{\urlprefix }}%
\providecommand \urlprefix  [0]{URL }%
\providecommand \Eprint [0]{\href }%
\providecommand \doibase [0]{http://dx.doi.org/}%
\providecommand \selectlanguage [0]{\@gobble}%
\providecommand \bibinfo  [0]{\@secondoftwo}%
\providecommand \bibfield  [0]{\@secondoftwo}%
\providecommand \translation [1]{[#1]}%
\providecommand \BibitemOpen [0]{}%
\providecommand \bibitemStop [0]{}%
\providecommand \bibitemNoStop [0]{.\EOS\space}%
\providecommand \EOS [0]{\spacefactor3000\relax}%
\providecommand \BibitemShut  [1]{\csname bibitem#1\endcsname}%
\let\auto@bib@innerbib\@empty
\bibitem [{\citenamefont {G{\"u}th}\ \emph {et~al.}(1982)\citenamefont
  {G{\"u}th}, \citenamefont {Schmittberger},\ and\ \citenamefont
  {Schwarze}}]{Guth1982An}%
  \BibitemOpen
  \bibfield  {author} {\bibinfo {author} {\bibfnamefont {W.}~\bibnamefont
  {G{\"u}th}}, \bibinfo {author} {\bibfnamefont {R.~W.}\ \bibnamefont
  {Schmittberger}}, \ and\ \bibinfo {author} {\bibfnamefont {B.}~\bibnamefont
  {Schwarze}},\ }\href {\doibase https://doi.org/10.1016/0167-2681(82)90011-7}
  {\bibfield  {journal} {\bibinfo  {journal} {Journal of Economic Behavior and
  Organization}\ }\textbf {\bibinfo {volume} {3}},\ \bibinfo {pages} {367}
  (\bibinfo {year} {1982})}\BibitemShut {NoStop}%
\bibitem [{\citenamefont {Solow}\ and\ \citenamefont
  {Simon}(1957)}]{Simon1957Models}%
  \BibitemOpen
  \bibfield  {author} {\bibinfo {author} {\bibfnamefont {R.~M.}\ \bibnamefont
  {Solow}}\ and\ \bibinfo {author} {\bibfnamefont {H.~A.}\ \bibnamefont
  {Simon}},\ }\href@noop {} {\emph {\bibinfo {title} {Models of Man: Social and
  Rational}}}\ (\bibinfo  {publisher} {Willey},\ \bibinfo {year}
  {1957})\BibitemShut {NoStop}%
\bibitem [{\citenamefont {Samuelson}\ and\ \citenamefont
  {Nordhaus}(2005)}]{Samuelson2005Economics}%
  \BibitemOpen
  \bibfield  {author} {\bibinfo {author} {\bibfnamefont {P.}~\bibnamefont
  {Samuelson}}\ and\ \bibinfo {author} {\bibfnamefont {W.}~\bibnamefont
  {Nordhaus}},\ }\href@noop {} {\emph {\bibinfo {title} {Economics (18th
  edition)}}}\ (\bibinfo  {publisher} {McGraw-Hil Education},\ \bibinfo {year}
  {2005})\BibitemShut {NoStop}%
\bibitem [{\citenamefont {Thaler}(1988)}]{Thaler1988Anomalies}%
  \BibitemOpen
  \bibfield  {author} {\bibinfo {author} {\bibfnamefont {R.~H.}\ \bibnamefont
  {Thaler}},\ }\href {\doibase 10.1257/jep.2.4.195} {\bibfield  {journal}
  {\bibinfo  {journal} {Journal of Economic Perspectives}\ }\textbf {\bibinfo
  {volume} {2}},\ \bibinfo {pages} {195} (\bibinfo {year} {1988})}\BibitemShut
  {NoStop}%
\bibitem [{\citenamefont {Bolton}\ and\ \citenamefont
  {Zwick}(1995)}]{Bolton1995Anonymity}%
  \BibitemOpen
  \bibfield  {author} {\bibinfo {author} {\bibfnamefont {G.~E.}\ \bibnamefont
  {Bolton}}\ and\ \bibinfo {author} {\bibfnamefont {R.}~\bibnamefont {Zwick}},\
  }\href {\doibase https://doi.org/10.1006/game.1995.1026} {\bibfield
  {journal} {\bibinfo  {journal} {Games and Economic Behavior}\ }\textbf
  {\bibinfo {volume} {10}},\ \bibinfo {pages} {95} (\bibinfo {year}
  {1995})}\BibitemShut {NoStop}%
\bibitem [{\citenamefont {Kagel}\ and\ \citenamefont
  {Roth}(1995)}]{Roth1995The}%
  \BibitemOpen
  \bibfield  {author} {\bibinfo {author} {\bibfnamefont {J.~H.}\ \bibnamefont
  {Kagel}}\ and\ \bibinfo {author} {\bibfnamefont {A.~E.}\ \bibnamefont
  {Roth}},\ }\href@noop {} {\emph {\bibinfo {title} {The Handbook of
  Experimental Economics}}}\ (\bibinfo  {publisher} {Princeton University
  Press, Princeton, NJ},\ \bibinfo {year} {1995})\BibitemShut {NoStop}%
\bibitem [{\citenamefont {Guth}\ and\ \citenamefont
  {Kocher}(2014)}]{Guth2014More}%
  \BibitemOpen
  \bibfield  {author} {\bibinfo {author} {\bibfnamefont {W.}~\bibnamefont
  {Guth}}\ and\ \bibinfo {author} {\bibfnamefont {M.~G.}\ \bibnamefont
  {Kocher}},\ }\href {\doibase https://doi.org/10.1016/j.jebo.2014.06.006}
  {\bibfield  {journal} {\bibinfo  {journal} {Journal of Economic Behavior and
  Organization}\ }\textbf {\bibinfo {volume} {108}},\ \bibinfo {pages} {396}
  (\bibinfo {year} {2014})}\BibitemShut {NoStop}%
\bibitem [{\citenamefont {Brosnan}\ and\ \citenamefont
  {de~Waal}(2003)}]{Brosnan2003Monkeys}%
  \BibitemOpen
  \bibfield  {author} {\bibinfo {author} {\bibfnamefont {S.~F.}\ \bibnamefont
  {Brosnan}}\ and\ \bibinfo {author} {\bibfnamefont {F.~B.~M.}\ \bibnamefont
  {de~Waal}},\ }\href {\doibase 10.1038/nature01963} {\bibfield  {journal}
  {\bibinfo  {journal} {Nature}\ }\textbf {\bibinfo {volume} {425}},\ \bibinfo
  {pages} {297} (\bibinfo {year} {2003})}\BibitemShut {NoStop}%
\bibitem [{\citenamefont {Proctor}\ \emph {et~al.}(2013)\citenamefont
  {Proctor}, \citenamefont {Williamson}, \citenamefont {de~Waal},\ and\
  \citenamefont {Brosnan}}]{Proctor2013Chimpanzees}%
  \BibitemOpen
  \bibfield  {author} {\bibinfo {author} {\bibfnamefont {D.}~\bibnamefont
  {Proctor}}, \bibinfo {author} {\bibfnamefont {R.}~\bibnamefont {Williamson}},
  \bibinfo {author} {\bibfnamefont {F.~B.~M.}\ \bibnamefont {de~Waal}}, \ and\
  \bibinfo {author} {\bibfnamefont {S.~F.}\ \bibnamefont {Brosnan}},\ }\href
  {\doibase 10.1073/pnas.1220806110} {\bibfield  {journal} {\bibinfo  {journal}
  {Proc. Nat. Acad. Sci. USA}\ }\textbf {\bibinfo {volume} {110}},\ \bibinfo
  {pages} {2070} (\bibinfo {year} {2013})}\BibitemShut {NoStop}%
\bibitem [{\citenamefont {Keith~Jensen}\ and\ \citenamefont
  {Tomasello}(2007)}]{Jensen2007Chimpanzees}%
  \BibitemOpen
  \bibfield  {author} {\bibinfo {author} {\bibfnamefont {J.~C.}\ \bibnamefont
  {Keith~Jensen}}\ and\ \bibinfo {author} {\bibfnamefont {M.}~\bibnamefont
  {Tomasello}},\ }\href {\doibase 10.1126/science.1145850} {\bibfield
  {journal} {\bibinfo  {journal} {Science}\ }\textbf {\bibinfo {volume}
  {318}},\ \bibinfo {pages} {107} (\bibinfo {year} {2007})}\BibitemShut
  {NoStop}%
\bibitem [{\citenamefont {Kirchsteiger}(1994)}]{Kirchsteiger1994The}%
  \BibitemOpen
  \bibfield  {author} {\bibinfo {author} {\bibfnamefont {G.}~\bibnamefont
  {Kirchsteiger}},\ }\href {\doibase
  https://doi.org/10.1016/0167-2681(94)90106-6} {\bibfield  {journal} {\bibinfo
   {journal} {Journal of Economic Behavior and Organization}\ }\textbf
  {\bibinfo {volume} {25}},\ \bibinfo {pages} {373} (\bibinfo {year}
  {1994})}\BibitemShut {NoStop}%
\bibitem [{\citenamefont {Bethwaite}\ and\ \citenamefont
  {Tompkinson}(1996)}]{Bethwaite1996The}%
  \BibitemOpen
  \bibfield  {author} {\bibinfo {author} {\bibfnamefont {J.}~\bibnamefont
  {Bethwaite}}\ and\ \bibinfo {author} {\bibfnamefont {P.}~\bibnamefont
  {Tompkinson}},\ }\href {\doibase
  https://doi.org/10.1016/0167-4870(96)00006-2} {\bibfield  {journal} {\bibinfo
   {journal} {Journal of Economic Psychology}\ }\textbf {\bibinfo {volume}
  {17}},\ \bibinfo {pages} {259} (\bibinfo {year} {1996})}\BibitemShut
  {NoStop}%
\bibitem [{\citenamefont {Fehr}\ and\ \citenamefont
  {Schmidt}(1999)}]{Fehr1999}%
  \BibitemOpen
  \bibfield  {author} {\bibinfo {author} {\bibfnamefont {E.}~\bibnamefont
  {Fehr}}\ and\ \bibinfo {author} {\bibfnamefont {K.}~\bibnamefont {Schmidt}},\
  }\href {\doibase 10} {\bibfield  {journal} {\bibinfo  {journal} {The
  Quarterly Journal of Economics}\ }\textbf {\bibinfo {volume} {114}},\
  \bibinfo {pages} {817} (\bibinfo {year} {1999})}\BibitemShut {NoStop}%
\bibitem [{\citenamefont {Debove}\ \emph {et~al.}(2016)\citenamefont {Debove},
  \citenamefont {Baumard},\ and\ \citenamefont {Andr{\'e}}}]{Debove2016Models}%
  \BibitemOpen
  \bibfield  {author} {\bibinfo {author} {\bibfnamefont {S.}~\bibnamefont
  {Debove}}, \bibinfo {author} {\bibfnamefont {N.}~\bibnamefont {Baumard}}, \
  and\ \bibinfo {author} {\bibfnamefont {J.-B.}\ \bibnamefont {Andr{\'e}}},\
  }\href {\doibase https://doi.org/10.1016/j.evolhumbehav.2016.01.001}
  {\bibfield  {journal} {\bibinfo  {journal} {Evolution and Human Behavior}\
  }\textbf {\bibinfo {volume} {37}},\ \bibinfo {pages} {245} (\bibinfo {year}
  {2016})}\BibitemShut {NoStop}%
\bibitem [{\citenamefont {Nowak}\ \emph {et~al.}(2000)\citenamefont {Nowak},
  \citenamefont {Page},\ and\ \citenamefont {Sigmund}}]{Nowak2000Fairness}%
  \BibitemOpen
  \bibfield  {author} {\bibinfo {author} {\bibfnamefont {M.~A.}\ \bibnamefont
  {Nowak}}, \bibinfo {author} {\bibfnamefont {K.~M.}\ \bibnamefont {Page}}, \
  and\ \bibinfo {author} {\bibfnamefont {K.}~\bibnamefont {Sigmund}},\ }\href
  {\doibase 10.1126/science.289.5485.1773} {\bibfield  {journal} {\bibinfo
  {journal} {Science}\ }\textbf {\bibinfo {volume} {289}},\ \bibinfo {pages}
  {1773} (\bibinfo {year} {2000})}\BibitemShut {NoStop}%
\bibitem [{\citenamefont {Andre}\ and\ \citenamefont
  {Baumard}(2011)}]{Andre2011Social}%
  \BibitemOpen
  \bibfield  {author} {\bibinfo {author} {\bibfnamefont {J.-B.}\ \bibnamefont
  {Andre}}\ and\ \bibinfo {author} {\bibfnamefont {N.}~\bibnamefont
  {Baumard}},\ }\href {\doibase https://doi.org/10.1016/j.jtbi.2011.07.031}
  {\bibfield  {journal} {\bibinfo  {journal} {Journal of Theoretical Biology}\
  }\textbf {\bibinfo {volume} {289}},\ \bibinfo {pages} {128} (\bibinfo {year}
  {2011})}\BibitemShut {NoStop}%
\bibitem [{\citenamefont {Binmore}\ and\ \citenamefont
  {Samuelson}(1994)}]{Binmore1994An}%
  \BibitemOpen
  \bibfield  {author} {\bibinfo {author} {\bibfnamefont {K.}~\bibnamefont
  {Binmore}}\ and\ \bibinfo {author} {\bibfnamefont {L.}~\bibnamefont
  {Samuelson}},\ }\href {\doibase 10.2307/40753015} {\bibfield  {journal}
  {\bibinfo  {journal} {Journal of Institutional and Theoretical Economics}\
  }\textbf {\bibinfo {volume} {150}},\ \bibinfo {pages} {45} (\bibinfo {year}
  {1994})}\BibitemShut {NoStop}%
\bibitem [{\citenamefont {Gale}\ \emph {et~al.}(1995)\citenamefont {Gale},
  \citenamefont {Binmore},\ and\ \citenamefont {Samuelson}}]{Gale1995Learning}%
  \BibitemOpen
  \bibfield  {author} {\bibinfo {author} {\bibfnamefont {J.}~\bibnamefont
  {Gale}}, \bibinfo {author} {\bibfnamefont {K.~G.}\ \bibnamefont {Binmore}}, \
  and\ \bibinfo {author} {\bibfnamefont {L.}~\bibnamefont {Samuelson}},\ }\href
  {\doibase https://doi.org/10.1016/S0899-8256(05)80017-X} {\bibfield
  {journal} {\bibinfo  {journal} {Games and Economic Behavior}\ }\textbf
  {\bibinfo {volume} {8}},\ \bibinfo {pages} {56} (\bibinfo {year}
  {1995})}\BibitemShut {NoStop}%
\bibitem [{\citenamefont {Sanchez}\ and\ \citenamefont
  {Cuesta}(2005)}]{Sanchez2005Altruism}%
  \BibitemOpen
  \bibfield  {author} {\bibinfo {author} {\bibfnamefont {A.}~\bibnamefont
  {Sanchez}}\ and\ \bibinfo {author} {\bibfnamefont {J.~A.}\ \bibnamefont
  {Cuesta}},\ }\href {\doibase https://doi.org/10.1016/j.jtbi.2005.01.006}
  {\bibfield  {journal} {\bibinfo  {journal} {Journal of Theoretical Biology}\
  }\textbf {\bibinfo {volume} {235}},\ \bibinfo {pages} {233} (\bibinfo {year}
  {2005})}\BibitemShut {NoStop}%
\bibitem [{\citenamefont {Page}\ and\ \citenamefont
  {Nowak}(2002)}]{Page2002Empathy}%
  \BibitemOpen
  \bibfield  {author} {\bibinfo {author} {\bibfnamefont {K.~M.}\ \bibnamefont
  {Page}}\ and\ \bibinfo {author} {\bibfnamefont {M.~A.}\ \bibnamefont
  {Nowak}},\ }\href {\doibase https://doi.org/10.1006/bulm.2002.0321}
  {\bibfield  {journal} {\bibinfo  {journal} {Bulletin of Mathematical
  Biology}\ }\textbf {\bibinfo {volume} {64}},\ \bibinfo {pages} {1101}
  (\bibinfo {year} {2002})}\BibitemShut {NoStop}%
\bibitem [{\citenamefont {Page}\ and\ \citenamefont
  {Sigmund}(2000)}]{Page2000The}%
  \BibitemOpen
  \bibfield  {author} {\bibinfo {author} {\bibfnamefont {K.~M.}\ \bibnamefont
  {Page}}\ and\ \bibinfo {author} {\bibfnamefont {N.~K.}\ \bibnamefont
  {Sigmund}},\ }\href {\doibase 10.1098/rspb.2000.1266} {\bibfield  {journal}
  {\bibinfo  {journal} {Proceedings Biological Sciences}\ }\textbf {\bibinfo
  {volume} {267}},\ \bibinfo {pages} {2177} (\bibinfo {year}
  {2000})}\BibitemShut {NoStop}%
\bibitem [{\citenamefont {Kuperman}\ and\ \citenamefont
  {Risau-Gusman}(2008)}]{Kuperman2008The}%
  \BibitemOpen
  \bibfield  {author} {\bibinfo {author} {\bibfnamefont {M.~N.}\ \bibnamefont
  {Kuperman}}\ and\ \bibinfo {author} {\bibfnamefont {S.}~\bibnamefont
  {Risau-Gusman}},\ }\href {\doibase 10.1140/epjb/e2008-00133-x} {\bibfield
  {journal} {\bibinfo  {journal} {European Physical Journal B}\ }\textbf
  {\bibinfo {volume} {62}},\ \bibinfo {pages} {233} (\bibinfo {year}
  {2008})}\BibitemShut {NoStop}%
\bibitem [{\citenamefont {Sinatra}\ \emph {et~al.}(2009)\citenamefont
  {Sinatra}, \citenamefont {Iranzo}, \citenamefont {G'omez-Gardenes},
  \citenamefont {Flor{\'i}a}, \citenamefont {Latora},\ and\ \citenamefont
  {Moreno}}]{Sinatra2009The}%
  \BibitemOpen
  \bibfield  {author} {\bibinfo {author} {\bibfnamefont {R.}~\bibnamefont
  {Sinatra}}, \bibinfo {author} {\bibfnamefont {J.}~\bibnamefont {Iranzo}},
  \bibinfo {author} {\bibfnamefont {J.}~\bibnamefont {G'omez-Gardenes}},
  \bibinfo {author} {\bibfnamefont {L.~M.}\ \bibnamefont {Flor{\'i}a}},
  \bibinfo {author} {\bibfnamefont {V.}~\bibnamefont {Latora}}, \ and\ \bibinfo
  {author} {\bibfnamefont {Y.}~\bibnamefont {Moreno}},\ }\href {\doibase
  10.1088/1742-5468/2009/09/P09012} {\bibfield  {journal} {\bibinfo  {journal}
  {Journal of Statistical Mechanics: Theory and Experiment}\ }\textbf {\bibinfo
  {volume} {58}},\ \bibinfo {pages} {09012} (\bibinfo {year}
  {2009})}\BibitemShut {NoStop}%
\bibitem [{\citenamefont {Iranzo}\ \emph {et~al.}(2011)\citenamefont {Iranzo},
  \citenamefont {Rom{\'a}n},\ and\ \citenamefont
  {S{\'a}nchez}}]{Iranzo2011The}%
  \BibitemOpen
  \bibfield  {author} {\bibinfo {author} {\bibfnamefont {J.}~\bibnamefont
  {Iranzo}}, \bibinfo {author} {\bibfnamefont {J.~M.}\ \bibnamefont
  {Rom{\'a}n}}, \ and\ \bibinfo {author} {\bibfnamefont {{\'A}.}~\bibnamefont
  {S{\'a}nchez}},\ }\href {\doibase https://doi.org/10.1016/j.jtbi.2011.02.020}
  {\bibfield  {journal} {\bibinfo  {journal} {Journal of Theoretical Biology}\
  }\textbf {\bibinfo {volume} {278}},\ \bibinfo {pages} {1} (\bibinfo {year}
  {2011})}\BibitemShut {NoStop}%
\bibitem [{\citenamefont {Colin F.~Camerer}\ and\ \citenamefont
  {Rabin}(2003)}]{Camerer2003Advances}%
  \BibitemOpen
  \bibfield  {author} {\bibinfo {author} {\bibfnamefont {G.~L.}\ \bibnamefont
  {Colin F.~Camerer}}\ and\ \bibinfo {author} {\bibfnamefont {M.}~\bibnamefont
  {Rabin}},\ }\href@noop {} {\emph {\bibinfo {title} {Advances in Behavioral
  Economics}}}\ (\bibinfo  {publisher} {Princeton University Press, Princeton,
  NJ},\ \bibinfo {year} {2003})\BibitemShut {NoStop}%
\bibitem [{\citenamefont {Loewenstein}\ and\ \citenamefont
  {Lerner}(2003)}]{Loewenstein2003The}%
  \BibitemOpen
  \bibfield  {author} {\bibinfo {author} {\bibfnamefont {G.}~\bibnamefont
  {Loewenstein}}\ and\ \bibinfo {author} {\bibfnamefont {J.~S.}\ \bibnamefont
  {Lerner}},\ }\href@noop {} {\emph {\bibinfo {title} {The Role of Affect in
  Decision Making}}}\ (\bibinfo  {publisher} {Oxford University Press,
  Oxford},\ \bibinfo {year} {2003})\BibitemShut {NoStop}%
\bibitem [{\citenamefont {Sanfey}\ \emph {et~al.}(2003)\citenamefont {Sanfey},
  \citenamefont {Rilling}, \citenamefont {Aronson}, \citenamefont {Nystrom},\
  and\ \citenamefont {Cohen}}]{Sanfey2003The}%
  \BibitemOpen
  \bibfield  {author} {\bibinfo {author} {\bibfnamefont {A.~G.}\ \bibnamefont
  {Sanfey}}, \bibinfo {author} {\bibfnamefont {J.~K.}\ \bibnamefont {Rilling}},
  \bibinfo {author} {\bibfnamefont {J.~A.}\ \bibnamefont {Aronson}}, \bibinfo
  {author} {\bibfnamefont {L.}~\bibnamefont {Nystrom}}, \ and\ \bibinfo
  {author} {\bibfnamefont {J.~D.}\ \bibnamefont {Cohen}},\ }\href {\doibase
  10.1126/science.1082976} {\bibfield  {journal} {\bibinfo  {journal}
  {Science}\ }\textbf {\bibinfo {volume} {300}},\ \bibinfo {pages} {1755}
  (\bibinfo {year} {2003})}\BibitemShut {NoStop}%
\bibitem [{\citenamefont {Rilling}\ and\ \citenamefont
  {Sanfey}(2011)}]{Rilling2011The}%
  \BibitemOpen
  \bibfield  {author} {\bibinfo {author} {\bibfnamefont {J.~K.}\ \bibnamefont
  {Rilling}}\ and\ \bibinfo {author} {\bibfnamefont {A.~G.}\ \bibnamefont
  {Sanfey}},\ }\href {\doibase 10.1146/annurev.psych.121208.131647} {\bibfield
  {journal} {\bibinfo  {journal} {Annual Review of Psychology}\ }\textbf
  {\bibinfo {volume} {62}},\ \bibinfo {pages} {23} (\bibinfo {year}
  {2011})}\BibitemShut {NoStop}%
\bibitem [{\citenamefont {Alexander}(2007)}]{2007The}%
  \BibitemOpen
  \bibfield  {author} {\bibinfo {author} {\bibfnamefont {J.~M.}\ \bibnamefont
  {Alexander}},\ }\href@noop {} {\emph {\bibinfo {title} {The Structural
  Evolution of Morality}}}\ (\bibinfo  {publisher} {Cambridge University
  Press},\ \bibinfo {year} {2007})\BibitemShut {NoStop}%
\bibitem [{\citenamefont {Frey}(2000)}]{2000The}%
  \BibitemOpen
  \bibfield  {author} {\bibinfo {author} {\bibfnamefont {B.~F.}\ \bibnamefont
  {Frey}},\ }\href {\doibase 10.1023/A:1006139124110} {\bibfield  {journal}
  {\bibinfo  {journal} {Journal of Business Ethics}\ }\textbf {\bibinfo
  {volume} {26}},\ \bibinfo {pages} {181} (\bibinfo {year} {2000})}\BibitemShut
  {NoStop}%
\bibitem [{\citenamefont {Boehm}(2012)}]{Boehm2012Moral}%
  \BibitemOpen
  \bibfield  {author} {\bibinfo {author} {\bibfnamefont {C.}~\bibnamefont
  {Boehm}},\ }\href@noop {} {\emph {\bibinfo {title} {Moral Origins: The
  Evolution of Virtue, Altruism, and Shame}}}\ (\bibinfo  {publisher} {Basic
  Books},\ \bibinfo {year} {2012})\BibitemShut {NoStop}%
\bibitem [{\citenamefont {Roca}\ \emph {et~al.}(2009)\citenamefont {Roca},
  \citenamefont {Cuesta},\ and\ \citenamefont
  {S{\'a}nchez}}]{Roca2009Evolutionary}%
  \BibitemOpen
  \bibfield  {author} {\bibinfo {author} {\bibfnamefont {C.~P.}\ \bibnamefont
  {Roca}}, \bibinfo {author} {\bibfnamefont {J.~A.}\ \bibnamefont {Cuesta}}, \
  and\ \bibinfo {author} {\bibfnamefont {A.}~\bibnamefont {S{\'a}nchez}},\
  }\href {\doibase 10.1016/j.plrev.2009.08.001} {\bibfield  {journal} {\bibinfo
   {journal} {Physics of Life Reviews}\ }\textbf {\bibinfo {volume} {6}},\
  \bibinfo {pages} {208} (\bibinfo {year} {2009})}\BibitemShut {NoStop}%
\bibitem [{\citenamefont {Fan}\ \emph {et~al.}(2020)\citenamefont {Fan},
  \citenamefont {Meng}, \citenamefont {Liu}, \citenamefont {Saberi},
  \citenamefont {Kurths},\ and\ \citenamefont {Nagler}}]{Fan2020Universal}%
  \BibitemOpen
  \bibfield  {author} {\bibinfo {author} {\bibfnamefont {J.}~\bibnamefont
  {Fan}}, \bibinfo {author} {\bibfnamefont {J.}~\bibnamefont {Meng}}, \bibinfo
  {author} {\bibfnamefont {Y.}~\bibnamefont {Liu}}, \bibinfo {author}
  {\bibfnamefont {A.~A.}\ \bibnamefont {Saberi}}, \bibinfo {author}
  {\bibfnamefont {J.}~\bibnamefont {Kurths}}, \ and\ \bibinfo {author}
  {\bibfnamefont {J.}~\bibnamefont {Nagler}},\ }\href {\doibase
  10.1038/s41567-019-0783-2} {\bibfield  {journal} {\bibinfo  {journal} {Nat.
  Phys.}\ }\textbf {\bibinfo {volume} {16}},\ \bibinfo {pages} {455} (\bibinfo
  {year} {2020})}\BibitemShut {NoStop}%
\bibitem [{\citenamefont {Stanley}(1987)}]{1987Introduction}%
  \BibitemOpen
  \bibfield  {author} {\bibinfo {author} {\bibfnamefont {H.~E.}\ \bibnamefont
  {Stanley}},\ }\href@noop {} {\emph {\bibinfo {title} {Introduction to Phase
  Transitions and Critical Phenomena}}}\ (\bibinfo  {publisher} {Oxford
  University Press},\ \bibinfo {year} {1987})\BibitemShut {NoStop}%
\bibitem [{\citenamefont {Taylor}\ and\ \citenamefont
  {Jonker}(1978)}]{Taylor1978Evolutionary}%
  \BibitemOpen
  \bibfield  {author} {\bibinfo {author} {\bibfnamefont {P.~D.}\ \bibnamefont
  {Taylor}}\ and\ \bibinfo {author} {\bibfnamefont {L.~B.}\ \bibnamefont
  {Jonker}},\ }\href {\doibase 10.1016/0025-5564(78)90077-9} {\bibfield
  {journal} {\bibinfo  {journal} {Mathematical Biosciences}\ }\textbf {\bibinfo
  {volume} {40}},\ \bibinfo {pages} {145} (\bibinfo {year} {1978})}\BibitemShut
  {NoStop}%
\bibitem [{\citenamefont {Kuznetsov}(2004)}]{Kuznetsov2004Elements}%
  \BibitemOpen
  \bibfield  {author} {\bibinfo {author} {\bibfnamefont {Y.~A.}\ \bibnamefont
  {Kuznetsov}},\ }\href@noop {} {\emph {\bibinfo {title} {Elements of applied
  bifurcation theory (3rd Edition)}}}\ (\bibinfo  {publisher} {Springer, New
  York, NY},\ \bibinfo {year} {2004})\BibitemShut {NoStop}%
\bibitem [{\citenamefont {Strogatz}(2015)}]{Strogatz2015Nonlinear}%
  \BibitemOpen
  \bibfield  {author} {\bibinfo {author} {\bibfnamefont {S.~H.}\ \bibnamefont
  {Strogatz}},\ }\href@noop {} {\emph {\bibinfo {title} {Nonlinear Dynamics and
  Chaos: With Applications to Physics, Biology, Chemistry, and Engineering (2nd
  Edition)}}}\ (\bibinfo  {publisher} {Westview Press},\ \bibinfo {year}
  {2015})\BibitemShut {NoStop}%
\bibitem [{\citenamefont {Watts}\ and\ \citenamefont
  {Strogatz}(1998)}]{1998Collective}%
  \BibitemOpen
  \bibfield  {author} {\bibinfo {author} {\bibfnamefont {D.~J.}\ \bibnamefont
  {Watts}}\ and\ \bibinfo {author} {\bibfnamefont {S.~H.}\ \bibnamefont
  {Strogatz}},\ }\href {\doibase 10.1038/30918} {\bibfield  {journal} {\bibinfo
   {journal} {Nature}\ }\textbf {\bibinfo {volume} {393}},\ \bibinfo {pages}
  {440} (\bibinfo {year} {1998})}\BibitemShut {NoStop}%
\bibitem [{\citenamefont {Bollob{\'a}s}(2001)}]{Bollobas2001random}%
  \BibitemOpen
  \bibfield  {author} {\bibinfo {author} {\bibfnamefont {B.}~\bibnamefont
  {Bollob{\'a}s}},\ }\href@noop {} {\emph {\bibinfo {title} {Random Graphs}}}\
  (\bibinfo  {publisher} {Cambridge University Press},\ \bibinfo {year}
  {2001})\BibitemShut {NoStop}%
\bibitem [{\citenamefont {Barab{\'a}si}\ and\ \citenamefont
  {Albert}(1999)}]{Barabasi1999Emergence}%
  \BibitemOpen
  \bibfield  {author} {\bibinfo {author} {\bibfnamefont {A.-L.}\ \bibnamefont
  {Barab{\'a}si}}\ and\ \bibinfo {author} {\bibfnamefont {R.}~\bibnamefont
  {Albert}},\ }\href {\doibase 10.1126/science.286.5439.509} {\bibfield
  {journal} {\bibinfo  {journal} {Science}\ }\textbf {\bibinfo {volume}
  {286}},\ \bibinfo {pages} {509} (\bibinfo {year} {1999})}\BibitemShut
  {NoStop}%
\bibitem [{\citenamefont {G{\'o}mez-Garde{\~n}es}\ \emph
  {et~al.}(2007)\citenamefont {G{\'o}mez-Garde{\~n}es}, \citenamefont
  {Moreno},\ and\ \citenamefont {Arenas}}]{GG2007Paths}%
  \BibitemOpen
  \bibfield  {author} {\bibinfo {author} {\bibfnamefont {J.}~\bibnamefont
  {G{\'o}mez-Garde{\~n}es}}, \bibinfo {author} {\bibfnamefont {Y.}~\bibnamefont
  {Moreno}}, \ and\ \bibinfo {author} {\bibfnamefont {A.}~\bibnamefont
  {Arenas}},\ }\href {\doibase 10.1103/PhysRevLett.98.034101} {\bibfield
  {journal} {\bibinfo  {journal} {Phys. Rev. Lett.}\ }\textbf {\bibinfo
  {volume} {98}},\ \bibinfo {pages} {034101} (\bibinfo {year}
  {2007})}\BibitemShut {NoStop}%
\bibitem [{\citenamefont {Santos}\ and\ \citenamefont
  {Pacheco}(2005)}]{Santos2005Scale-Free}%
  \BibitemOpen
  \bibfield  {author} {\bibinfo {author} {\bibfnamefont {F.~C.}\ \bibnamefont
  {Santos}}\ and\ \bibinfo {author} {\bibfnamefont {J.~M.}\ \bibnamefont
  {Pacheco}},\ }\href {\doibase 10.1103/PhysRevLett.95.098104} {\bibfield
  {journal} {\bibinfo  {journal} {Phys. Rev. Lett.}\ }\textbf {\bibinfo
  {volume} {95}},\ \bibinfo {pages} {098104} (\bibinfo {year}
  {2005})}\BibitemShut {NoStop}%
\bibitem [{\citenamefont {Szab{\'o}}\ and\ \citenamefont {T{\H
  o}ke}(1998)}]{Szabo1998Evolutionary}%
  \BibitemOpen
  \bibfield  {author} {\bibinfo {author} {\bibfnamefont {G.}~\bibnamefont
  {Szab{\'o}}}\ and\ \bibinfo {author} {\bibfnamefont {C.}~\bibnamefont {T{\H
  o}ke}},\ }\href {\doibase 10.1103/PhysRevE.58.69} {\bibfield  {journal}
  {\bibinfo  {journal} {Phys. Rev. E}\ }\textbf {\bibinfo {volume} {58}},\
  \bibinfo {pages} {69} (\bibinfo {year} {1998})}\BibitemShut {NoStop}%
\bibitem [{\citenamefont {Brenner}\ and\ \citenamefont
  {Vriend}(2006)}]{Brenner2006On}%
  \BibitemOpen
  \bibfield  {author} {\bibinfo {author} {\bibfnamefont {T.}~\bibnamefont
  {Brenner}}\ and\ \bibinfo {author} {\bibfnamefont {N.~J.}\ \bibnamefont
  {Vriend}},\ }\href {\doibase https://doi.org/10.1016/j.jebo.2004.07.014}
  {\bibfield  {journal} {\bibinfo  {journal} {Journal of Economic Behavior and
  Organization}\ }\textbf {\bibinfo {volume} {61}},\ \bibinfo {pages} {617}
  (\bibinfo {year} {2006})}\BibitemShut {NoStop}%
\bibitem [{\citenamefont {Xiong}\ \emph {et~al.}(2014)\citenamefont {Xiong},
  \citenamefont {Fu},\ and\ \citenamefont {Wang}}]{Xiong2014Emergence}%
  \BibitemOpen
  \bibfield  {author} {\bibinfo {author} {\bibfnamefont {W.}~\bibnamefont
  {Xiong}}, \bibinfo {author} {\bibfnamefont {H.}~\bibnamefont {Fu}}, \ and\
  \bibinfo {author} {\bibfnamefont {Y.}~\bibnamefont {Wang}},\ }\href {\doibase
  https://doi.org/10.1007/978-3-319-00912-4_9} {\emph {\bibinfo {title}
  {Emergence of Fair Offers in Ultimatum Game. In: Leitner S., Wall F. (eds)
  Artificial Economics and Self Organization. Lecture Notes in Economics and
  Mathematical Systems}}}\ (\bibinfo  {publisher} {Springer Press},\ \bibinfo
  {year} {2014})\BibitemShut {NoStop}%
\end{thebibliography}%
\end{document}